\documentclass{emulateapj}
\usepackage{amsmath,amssymb,gensymb,times,graphics,morefloats,color}
\usepackage{afterpage, bm}
\usepackage{natbib,hyperref}
\def\apj{ApJ}

\def\kepler{\emph{Kepler}}
 
\def\btsettl{BT-Settl}
\def\spex{SpeX}

\def\sun{\odot}

\def\lsun{L_{\odot}}
\def\rsun{R_{\odot}}
\def\teff{T_\mathrm{eff}}
\def\feh{\mathrm{[Fe/H]}}
\def\logg{\log g}
\def\dex{\mathrm{\ dex}}

\def\mk{\mathrm{M}_\mathrm{K}}

\def\lbol{L_\mathrm{bol}}
\def\fbol{F_\mathrm{bol}}
\def\rearth{R_\Earth}

\def\hind{\mathrm{H}_2\mathrm{O\mbox{-}K2}}

\def\H{$H$}
\def\K{$K$}
\def\J{$J$}

\def\tilde{\raise.17ex\hbox{$\scriptstyle\mathtt{\sim}$}}

\def\nlines{26}

\shorttitle{M dwarf temperatures and radii}
\shortauthors{Newton et al.}

\begin{document} 
\title{An empirical calibration to estimate cool dwarf fundamental parameters from \H-band spectra}
\author{Elisabeth R. Newton\altaffilmark{1}, David Charbonneau\altaffilmark{1}, Jonathan Irwin\altaffilmark{1}, Andrew W. Mann\altaffilmark{2}}
\altaffiltext{1}{Harvard-Smithsonian Center for Astrophysics, 60 Garden Street, Cambridge, MA 02138, USA}
\altaffiltext{2}{Astronomy Department, University of Texas at Austin, Austin, TX 78712, USA}

\begin{abstract}

Interferometric radius measurements provide a direct probe of the fundamental parameters of M dwarfs. However, interferometry is within reach for only a limited sample of nearby, bright stars. We use interferometrically-measured radii, bolometric luminosities, and effective temperatures to develop new empirical calibrations based on low-resolution, near-infrared spectra. We find that \H-band Mg and Al spectral features are good tracers of stellar properties, and derive functions that relate effective temperature, radius and log luminosity to these features. The standard deviations in the residuals of our best fits are, respectively, $73$K, $0.027\rsun$, and $0.049$ dex (an $11\%$ error on luminosity). Our calibrations are valid from mid-K to mid-M dwarf stars, roughly corresponding to temperatures between $3100$ and $4800$K.
We apply our \H-band calibrations to M dwarfs targeted by the MEarth transiting planet survey and to the cool \kepler\ Objects of Interest (KOIs). We present spectral measurements and estimated stellar parameters for these stars. Parallaxes are also available for many of the MEarth targets, allowing us to independently validate our calibrations by demonstrating a clear relationship between our inferred parameters and the stars' absolute \K\ magnitudes. We identify objects with magnitudes too bright for their inferred luminosities as candidate multiple systems. We also use our estimated luminosities to address the applicability of near-infrared metallicity calibrations to mid and late M dwarfs.
The temperatures we infer for the KOIs agree remarkably well with those from the literature; however, our stellar radii are systematically larger than those presented in previous works that derive radii from model isochrones. This results in a mean planet radius that is $15\%$ larger than one would infer using the stellar properties from recent catalogs. Our results confirm the derived parameters from previous in-depth studies of KOIs 961 (\kepler-42), 254 (\kepler-45), and 571 (\kepler-186), the latter of which hosts a rocky planet orbiting in its star's habitable zone.
\end{abstract}

\section{Introduction}

The characterization of planets is often limited by our knowledge of the stars they orbit. Stellar characterization is not yet as reliable for cooler dwarfs as it is for F, G and K dwarfs, for which fundamental stellar parameters can be determined with reasonable precision and accuracy \citep[e.g.][]{Valenti2005}.
The interpretation of observables for cooler dwarfs is complicated by uncertain sources of opacity and the appearance of complex molecules in their atmospheres, and by discrepancies between their observed and theoretical properties. For this reason, empirical calibrations remain an important component of our understanding of M dwarfs.

Empirically-derived relations between basic stellar parameters provide the basis for determining the physical properties of field M dwarfs. For stars with parallaxes, masses can be estimated using the mass to absolute \K-band magnitude relation of \citet[]{DelfosseX.2000}, and radii can then be calculated using a mass-radius relation \citep[e.g.,][]{Bayless2006,Boyajian2012}. The mass-magnitude and mass-radius relations are determined from double-lined eclipsing binaries and have precisions of about $10\%$. \citet{Johnson2012} circumvented the lack of a distance for \kepler\ Object of Interest (KOI) 254 by constraining the mass and radius of this star using four separate empirical relationships between photometric quantities, mass, radius, and $\feh$. Another technique that can be used is the infrared flux method \citep[IRFM;][]{Blackwell1977,Blackwell1979,Blackwell1980}, which was extended to M dwarfs by \citet{Casagrande2008}. The IRFM uses the ratio of infrared to bolometric flux and a grid of stellar models to determine the effective temperature ($\teff$) and bolometric flux ($\fbol$) of a star. 

Planet surveys have driven many authors to investigate methods to estimate the stellar parameters of cool dwarfs without parallaxes, which are predominantly based on fitting observations to grids of stellar models. This has proved particularly fruitful for M dwarfs targeted by the \kepler\ survey, some of which host confirmed or candidate planets -- including several likely to be Earth-sized and in their stars' habitable zone. \citet[]{Dressing2013} matched observed colors to Dartmouth stellar isochrones \citep[]{Dotter2008a, Feiden2011}. \citet{Mann2012} and \citet{Mann2013a} matched optical spectra to synthetic \btsettl\ and PHOENIX spectra, respectively. \citet[]{Muirhead2012a} and \citet{Muirhead2014} used moderate resolution $K$-band spectra to determine $\teff$ and metallicity for $103$ cool KOIs, which they interpolated onto Dartmouth isochrones to infer the stars' radii.

The methods described above rely on relationships derived from binary stars or on matching observed properties to stellar models. While there is reasonable agreement between predicted and observed masses and radii (or luminosities), the observed radii of M dwarfs at a given $\teff$ are larger than expected from models. This was first noted for binary stars \citep[e.g.][]{Popper1997,Torres2002,Berger2006,LopezMorales2007,Chabrier2007}, but has been demonstrated in stars with interferometric radii measurements \citep[e.g.][]{Berger2006,Boyajian2012}. Additionally, if magnetic activity is responsible for the inflated radii, as discussed in \citet{LopezMorales2005}, \citet{Ribas2006}, \citet{LopezMorales2007}, and \citet{Chabrier2007}, a mass-radius-\emph{activity} relationship would be required to accurately determine the radii of field dwarfs. 
On the theoretical side, synthetic spectra -- though they have been improved significantly over the last two decades -- suffer from incomplete line lists for the molecules that blanket the spectra, particularly for TiO, VO, metal hydrides, and water vapor \citep[e.g.][]{Valenti1998,Allard2000,Leggett2000,Bean2006,Onehag2012,Rajpurohit2013a}. 

The use of proxies -- stars with directly measured parameters that are similar to the star of interest -- enable to parameters of a field dwarf to be inferred with limited reliance on stellar models. \citet{Muirhead2012} used Barnard's star to anchor stellar models to direct measurements and inferred the properties of a similar M dwarf, KOI 961/\kepler-42. Building on the method used by \citet{Muirhead2012}, \citet{Ballard2013} used four stars with interferometric radii as proxies to infer the stellar properties of \kepler-61. As discussed by \citet{Ballard2013}, the use of proxies fills a particularly crucial niche: while the temperature-sensitive index used by \citet{Muirhead2012a,Muirhead2012,Muirhead2014} saturates $\teff>3800$K, the \kepler\ sample is rich in planets orbiting early M and late K dwarfs. \citet{Ballard2013} did not use models in their derivation of stellar parameters, instead directly using the interferometric radii. However, their method assumes that the properties of \kepler-61 match those of stars with the same spectral type, and M dwarf spectral types represent a coarse binning in stellar properties. It has also been shown that late-type dwarfs may have different spectral types in the optical and infrared \citep[e.g.][]{Rojas-Ayala2012,Pecaut2013,Newton2014}.

The idea of proxies naturally extends to the identification of empirical tracers of stellar parameters, and the development of such tracers is the motivation for this work. We develop methods to estimate stellar temperatures, radii, and luminosities that do not require parallaxes or stellar models, that are precise, accurate and extensible, and that can be applied to the large body of moderate-resolution near-infrared (NIR) spectra presently available. This idea was also used by \citet{Mann2013a}, who used the sample of stars with interferometric measurements to derive index-based calibrations for $\teff$ in the visible, $J$, $H$, and $K$ bands (although they use model-fitting for the stellar parameters reported therein). Our approach is also based on spectra of the interferometric sample but differs from their work in several ways. We utilize the NIR, which is not as strongly blanketed as the optical by molecular bands that are often sensitive to metallicity. Instead of spectral indices, we use equivalent widths (EWs), which do not require flux-calibrated spectra and are less sensitive to instrument characterization and atmospheric dispersion. Due to ongoing discussions in the literature on measurements of $\teff$, we opt to directly calibrate relations for $\teff$, radius, and luminosity, rather than inferring one property and using additional relations to determine the others. Finally, we have the benefit of three new interferometric measurements from \citet{vonBraun2014}, and include Gl 725B, which was excluded from the analysis in \citet[][]{Mann2013a}, in our radius calibration.

In this paper, we present purely empirical relations between the EWs of NIR features and a cool dwarf's $\teff$, radius, and luminosity. These calibrations are based on what we refer to as the interferometric sample: the set of stars with interferometrically-measured angular diameters and for which bolometric fluxes and distances have been measured, thereby allowing their effective temperatures and luminosities to be determined. We discuss our observations and measurements in \S\ref{Sec:observations}. In \S\ref{Sec:relations}, we discuss the behavior of \H-band spectral features and present our new calibrations, including discussion of systematic errors. 
We then apply our calibrations to MEarth M dwarfs in \S\ref{Sec:mearth} and consider the behavior of inferred stellar parameters with absolute \K\ magnitude. We also address the applicability of NIR metallicity calibrations from \citet{Mann2013}, \citet{Newton2014}, and \citet{Mann2014} to mid and late M dwarfs. We consider the cool KOIs in \S\ref{Sec:kepler}, and compare our stellar and planetary parameters to those derived using other techniques. In \S\ref{Sec:summary}, we summarize our findings.

\section{Observations and measurements}\label{Sec:observations}

Our observations and data reduction were carried out as discussed in \citet{Newton2014}, which we summarize briefly here. We used the \spex\ instrument on IRTF with the $0.3\times15\arcsec$ slit, yielding spectra from $0.8-2.4\micron$ with $R=2000$.  Four observations were acquired of each object, with two observations at each of two nod positions. Telluric standards were observed directly before or after observations of the interferometry stars used in this work. Flats (using an internal quartz lamp) and wavelength calibrations (using internal Thorium-Argon lamps) were taken throughout the night, at roughly one hour intervals or after large slews. 

We reduced our data using \texttt{Spextool} \citep[]{Cushing2004} and used \texttt{xtellcor} to perform telluric corrections \citep[]{Vacca2003}. We determined absolute radial velocities using the method described in \citet{Newton2014}, using telluric features to provide an absolute wavelength calibration. We then shifted the spectra to rest wavelengths. 

To measure EWs, we defined a wavelength range for the feature and nearby continuum regions on either side. We oversampled the data and numerically integrated the flux within the feature. We estimated errors on the radial velocities and equivalent widths by creating 50 realizations of our data, taking into account correlated noise for the high S/N observations of the interferometry and MEarth samples. We measured the EWs of \nlines\ spectral features and 13 spectral indices in the NIR, primarily the same lines as examined by \citet{Newton2014}. Our \H-band features are included in Table \ref{Tab:lines}. We measure the temperature-sensitive indices used in \citet{Mann2013a} and the $\hind$ index \citep{Rojas-Ayala2012}. Wavelengths are given in vacuum.

\begin{deluxetable}{l r r r r r r}
\tablecaption{\label{Tab:lines}\H-band spectral features}
\tablecolumns{7}
\tablehead{ \colhead{Feature} & \multicolumn{2}{c}{Feature window} & \multicolumn{2}{c}{Blue continuum} & \multicolumn{2}{c}{Red continuum} \\
& \multicolumn{2}{c}{\micron} & \multicolumn{2}{c}{\micron} & \multicolumn{2}{c}{\micron}}
\startdata
Mg ($1.48\micron$) & 1.4872 & 1.4892 & 1.4790 & 1.4850 & 1.4900 & 1.4950  \\
Mg ($1.50\micron$) & 1.5020 & 1.5060 & 1.4965 & 1.5000 & 1.5070 & 1.5120  \\
K ($1.51\micron$) & 1.5160 & 1.5180 & 1.5105 & 1.5135 & 1.5185 & 1.5215  \\
Mg ($1.57\micron$) & 1.5740 & 1.5780 & 1.5640 & 1.5680 & 1.5785 & 1.5815  \\
Si ($1.58\micron$) & 1.5880 & 1.5925 & 1.5845 & 1.5875 & 1.5925 & 1.5955  \\
CO ($1.61\micron$) & 1.6190 & 1.6220 & 1.6120 & 1.6150 & 1.6265 & 1.6295  \\
CO ($1.62\micron$)  & 1.6245 & 1.6265 & 1.6120 & 1.6150 & 1.6265 & 1.6295  \\
Al\mbox{-}a ($1.67\micron$)  & 1.6715 & 1.6735 & 1.6550 & 1.6650 & 1.6780 & 1.6820  \\
Al\mbox{-}b ($1.67\micron$)  & 1.6745 & 1.6775 & 1.6550 & 1.6650 & 1.6780 & 1.6820  \\
Mg ($1.71\micron$) & 1.7100 & 1.7125 & 1.7025 & 1.7055 & 1.7130 & 1.7160  \\
\enddata
\tablecomments{All wavelengths are given in vacuum.}
\end{deluxetable}

\subsection{Stars with interferometric measurements}\label{Sec:intstars}

Our calibration sample comprises 25 stars with interferometrically-measured radii. We preferentially use the interferometric measurements from \citet{Boyajian2012} and incorporate the measurements they collected from the literature. Where more than one such literature measurement is available, we use the weighted average. The literature sources for measurements are: \citet{Sgransan2003, Boyajian2008, Demory2009, Boyajian2012} and \citet{vonBraun2012}. We also include newer measurements from \citet{vonBraun2014}.

We observed 20 M dwarfs with interferometrically-measured radii from the sample described above. We supplement this sample with the five remaining: spectra for Gl 581 and Gl 892 are available in the IRTF spectral library \citep{Cushing2005, Rayner2009}\footnote{http://irtfweb.ifa.hawaii.edu/{\tilde}spex/IRTF\_Spectral\_Library/} and we include spectra of Gl 876, Gl 649, and Gl 176 that were observed by \citet{Mann2013a}. We cross-correlated these spectra with the RV standard we use in the rest of our work in order to assure agreement between our wavelength calibrations.

The bolometric fluxes and luminosities ($\fbol$ and $\lbol$) for the stars from \citet{Boyajian2012} were re-measured using spectra and photometry by \citet{Mann2013a}. To measure $\fbol$, \citet{Boyajian2012} use multicolor photometry to select best-fitting template spectra from the \citet{Pickles1998} catalog, which they extrapolate to the NIR using photometry. \citet{Mann2013a} demonstrated that the fluxes extrapolated beyond $1.1\micron$ do not match their observed spectra and instead used optical and infrared spectra that they obtained for each star. They use models to cover gaps in the spectra and assume Wein's and Rayleigh-Jeans' laws at shorter and longer wavelengths than covered by their spectra. They then adjust the overall flux level using a correction factor calculated by comparing photometric fluxes to the fluxes measured from the spectra. \citet{Mann2013a} measure $\teff$ systematically higher than \citet{Boyajian2012} by $1\%$. 

We use the updated parameters from \citet{Mann2013a} in this work, but note that the radii measured are insensitive to these changes in temperature. We apply the method from \citet{Mann2013a}, described above, to the three new objects from \citet{vonBraun2014} that we include in this paper. We provide updated parameters for Gl 176, Gl 649, and Gl 876 in Table \ref{Tab:update}. We also use the $\feh$ calibration from \citet{Mann2013} to estimate the iron abundances of these stars (see \S\ref{Sec:metallicities} for discussion of NIR metallicity calibrations).

Included in the table are the reduced $\chi^2$ ($\chi^2_{red}$) of the photometric corrections applied to each spectrum. Large $\chi^2_{red}$ indicate that errors in the photometry or spectrum are underestimated or that there are systematic offsets between the two. The $\chi^2_{red}$ for the three stars presented here are typical of those reported by \citet{Mann2013a}.

\begin{deluxetable*}{l r r r r r r}
\tablecaption{\label{Tab:update}Updated parameters for new interferometry stars}
\tablecolumns{7}
\tablehead{ 
	\colhead{Object} & 
	\colhead{$\fbol$} & 
	\colhead{$\teff$} &
	\colhead{$\lbol$} & 
	\colhead{$R$} &
	\colhead{$\chi^2_{red}$\tablenotemark{a}}   & 
	\colhead{$\feh$\tablenotemark{b}} \\
	& 
	\colhead{(ergs/s/cm$^2\times10^{-8}$)}& 
	\colhead{(K)} &
	\colhead{($\lsun$)} &
	\colhead{($\rsun$)} &
	&
	\colhead{(dex)}
	}
\startdata
Gl 176 & $1.2544\pm0.0099$ & $3701\pm90$ & $0.03352\pm0.00027$ &$ 0.4525\pm0.0221$ & 0.36 & $+0.17$\\
Gl 649 & $1.3171\pm0.0071$ & $3604\pm46$ & $0.04379\pm0.00023$ & $0.5387\pm0.0157$ & 1.94 & $+0.05$ \\
Gl 876 & $1.8863\pm0.0115$ & $3176\pm20$ & $0.01290\pm0.00008$ & $0.3761\pm0.0059$ & 0.78 & $+0.35$
\enddata
\tablenotetext{a}{$\chi^2_{red}$ describes the goodness of fit as described in the text and in \citet{Mann2013a}.}
\tablenotetext{b}{$\feh$ estimated from Equation 16 from \citet{Mann2013}.}
\end{deluxetable*}

\citet{Mann2013a} excluded Gl 725B (GJ 725B in their work) due to its atypically large $\chi^2_{red}$ (7.9). This indicates a large disagreement between this star's photometry and spectrum, which affects its $\fbol$ in an unknown manner. Therefore, we also exclude Gl 725B from our analysis of $\teff$ and $\lbol$, both of which are subject to this uncertainty. However, we include Gl 725B in our analysis of radius: while changing a star's temperature affects the interferometrically-measured radius through the adopted limb darkening model, the effect is small \citep[see][]{Boyajian2012}. Finally, we note that while Gl 725B has an unusually cool temperature given its radius, Gl 876 has similar properties; Gl 876 has $\chi^2_{red}=0.78$, typical of the values in \citet{Mann2013a}, and is not excluded from any part of our analysis.

\subsection{MEarth M dwarfs}

\citet{Newton2014} obtained spectra of $447$ mid to late M dwarfs in the solar neighborhood that are targets of the MEarth transiting planet survey \citep{Berta2012, Irwin2014}. Half of these M dwarfs had previously published parallaxes, which were compiled in \citet{Newton2014}.  \citet{Dittmann2014} measured parallaxes for most of these stars from MEarth astrometry; with the parallaxes from \citet{Dittmann2014}, we have distances to $388$ of these M dwarfs. 

We note that LSPM J0035+5241S is improperly matched in \citet{Newton2014}, who identified it as LSPM J0035+5241N. It is identified properly in this work as LSPM J0035+5241S. 

\subsection{Cool KOIs}\label{Sec:koiselection}

\citet{Muirhead2012a} and \citet{Muirhead2014} presented \H- and \K-band spectra for 103 KOIs with $r-J>2$, which implies they have temperatures below $4000$K. They obtained spectra using the TripleSpec instrument on the Palomar 200-inch (5.1m) Hale telescope \citep{Herter2008}, which simultaneously obtains $R=2700$ spectra in \J-, \H-, and \K-band. We convolved the spectra with a Gaussian of fixed width to degrade the resolution to that of IRTF. We then cross-correlated each spectrum with that of our RV standard to place them on the same wavelength calibration as our observations.

Measurements of EWs should not depend on the resolution of the spectrograph. However, most lines in the NIR spectra of a cool dwarf are not free of contaminating features at moderate resolution. We found systematic differences of $0.1$\AA\ between the EWs of some features measured at $R=2000$ and at $R=2700$, so we stress the importance of accounting for even moderate differences in spectral resolution. 

While the subtraction of sky emission lines in IRTF spectra is very robust, it is more difficult for TripleSpec spectra: because of the tilt of the slit, the removal of sky background requires that illumination be very well characterized. Sky emission features persist in some of the KOI spectra and contaminate many spectral features of interest, necessitating the removal of the affected spectra from our analysis. We first identified contaminated spectra by eye, then developed the following quantitative method to remove spectra based on scatter between the EWs of the components of doublets.

We compared the EWs of the two components of the Mg doublet at $1.50\micron$, of the \ion{Mg}{1} doublet at $1.57\micron$, and of the \ion{Al}{1} doublet at $1.67\micron$. One component in each of the $1.50\micron$ Mg doublet and the Al doublet is contaminated by sky emission. Neither component is contaminated in the $1.57\micron$ Mg doublet. We found that in the spectra of the MEarth and interferometry stars, for which sky emission is not prevalent, the EWs of the doublet components are linearly correlated with little scatter. For the $1.57\micron$ Mg doublet, which is free of sky emission, there is also little scatter for the KOIs. However, a fraction of the KOIs show inconsistencies in the EWs of the components of the contaminated  doublets. The stars that show scatter tend to be those that we determined by eye to contain sky emission. 

For each of the two doublets contaminated by sky emission, we used the interferometric sample to determine a linear correlation between the doublet components. Objects deviating from the best-fitting line by more than $0.75$\AA\ for Mg at $1.50\micron$ or $0.5$\AA\ for Al at $1.67\micron$ were discarded. These limits replicate the contamination we determined by eye and the amount of scatter expected based on observations of the MEarth and interferometry stars. Our final sample of KOIs includes 66 stars. The number of spectra excluded by our cuts does not depend sensitively on the limits we adopt.

\section{Inferring stellar parameters}\label{Sec:relations}

\begin{figure}
\includegraphics[width=\linewidth]{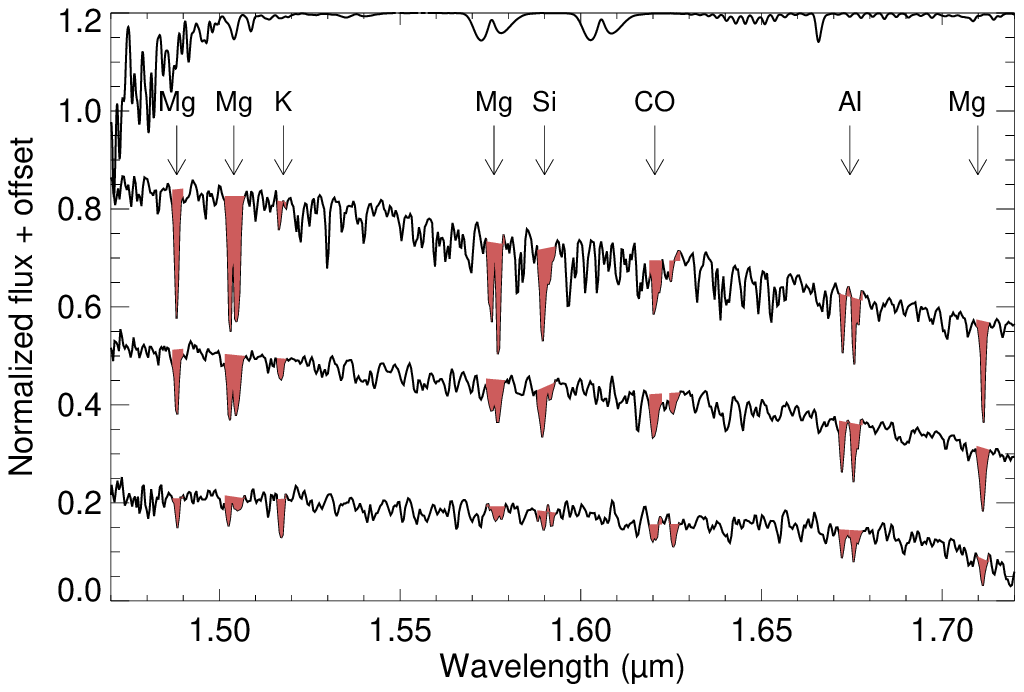}
\caption{Representative \H-band spectra of cool dwarfs spanning the temperature range of our calibration. From the top down: a K4V from the IRTF spectral library, an M2V composite spectrum, and an M5V composite spectrum. The composite spectra are from \citet{Newton2014}. We indicate the EW measurements with red shading and the elements that dominate the major absorption features. Along the top, we show a typical atmospheric transmission spectrum for Mauna Kea, from \citet{Lord1992}.
\label{Fig:spectra}}
\end{figure}

The interferometric sample, which we use to calibrate our empirical relationships, consists of 25 cool stars with radii directly measured using interferometry. We discussed observations of these stars in \S\ref{Sec:intstars}. Typical \H-band spectra, with our EW measurements indicated, are shown in Figure \ref{Fig:spectra}.

\subsection{Behavior of \H-band spectral features as a function of effective temperature}

\begin{figure*}
\includegraphics[width=\linewidth]{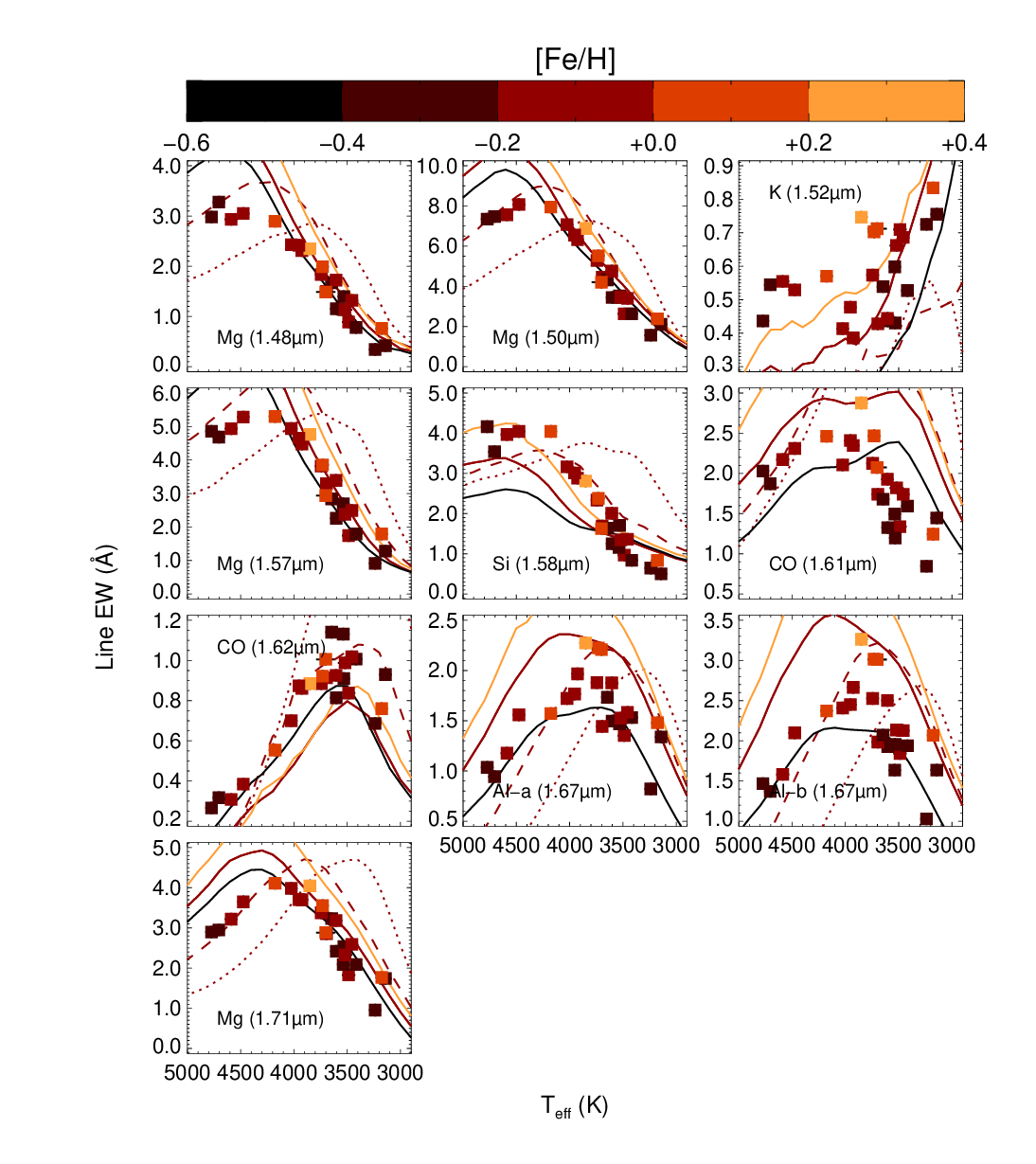}
\caption{EWs of \H-band spectral features plotted against measured $\teff$ for stars in our interferometric sample. We use the updated temperatures from \citet{Mann2013a} and adopt iron abundances using the \citet{Mann2013} calibration. The EWs measured from \btsettl\ model spectra \citep{Allard2014} are also plotted. Solid lines are $\logg=5$, dashed lines for $\logg=4.5$, and dotted lines for $\logg=4$, all with solar metallicity. For $\logg=5$, we show three metallicities: $-0.5$ dex (black), $+0.0$ dex (red), and $+0.3$ dex (orange).
\label{Fig:teff_line}}
\end{figure*}

\begin{figure*}
\includegraphics[width=\linewidth]{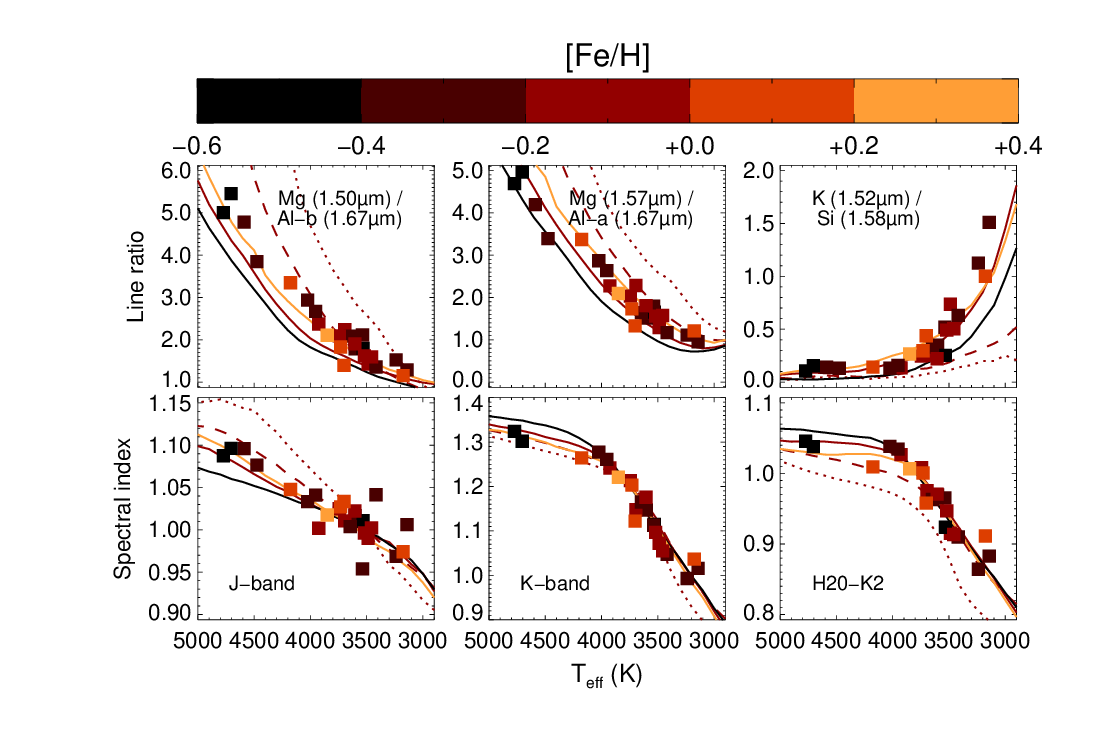}
\caption{Same as Figure \ref{Fig:teff_line} but plotting select EW ratios and spectral indices against $\teff$, using the updated temperatures from \citet{Mann2013a}.
\label{Fig:teff_feat}}
\end{figure*}

In Figures \ref{Fig:teff_line}-\ref{Fig:teff_feat}, we plot selected EWs, EW ratios, and spectral indices for the interferometric calibration sample against the measured $\teff$. Included in these plots are all \H-band EW measurements, the two EW ratios used in our final analysis (each of which is \ion{Mg}{1} / \ion{Al}{1}, involving different atomic transitions), and one EW ratio that we found to have a strong temperature dependence for mid M dwarfs. This last EW ratio, \ion{K}{1} / \ion{Si}{1} , may be useful for studies focusing on later objects. The data points in these figures are colored by their $\feh$, using the iron abundance calibration from \citet{Mann2013}.

We measured the same spectral features in \btsettl\ model spectra \citep[][available online\footnote{http://phoenix.ens-lyon.fr/Grids/BT-Settl/}]{Allard2014} as we do in our observed spectra. We used models based on the \citet{Asplund2009} solar abundances and degraded their resolution to that of \spex\ by convolving the spectra with a fixed-width Gaussian. We then measured EWs numerically using the same technique that we used with our observed spectra. Figures \ref{Fig:teff_line}-\ref{Fig:teff_feat} include the EWs, EW ratios, and spectral features that we measured for a suite of synthetic spectra. We show temperature tracks for three surface gravities, using solar metallicity: $\log g=5$, $4.5$, and $4$. Late dwarfs have long main sequence lifetimes, so age does not strongly affect $\log g$. For $\log g = 5$, we also show temperature tracks for two non-solar metallicities, $\feh = -0.5$ and $\feh = +0.3$, which roughly spans the metallicity range of the interferometric sample. The synthetic spectra with $\feh=-0.5$ have an alpha enhancement of $+0.2$; the remainder have no alpha enhancement. The EW measurements of M dwarfs should most closely match the theoretical $\log g = 5$ temperature track. The surface gravities of the late K dwarfs ($\teff>4000$) decrease with increasing temperature, reaching $\logg = 4.5$ for the hottest star in our sample. Observed EWs for the late K dwarfs are therefore expected to most closely match the temperature track with that surface gravity.

Spectral features that show the smallest metallicity dependence show the most agreement between the EWs measured in the synthetic spectra and in our observed spectra. The \ion{Mg}{1} EWs show the best agreement (though the slight metallicity dependence in the models is not evident in our data). For the M dwarfs, the EWs closely follow the temperature tracks for stars with $\log g = 5$, as expected. For the four K dwarfs, the EWs drop due to the lower surface gravities of these objects. For the \ion{Mg}{1} features at $1.50\micron$ and $1.71\micron$, the EWs of K dwarfs closely follow the temperature tracks for $\log g = 4.5$; however for those at $1.48\micron$ and $1.57\micron$, the EWs are smaller than expected for stars with their temperatures and surface gravities. The EWs of the \ion{Si}{1} feature at $1.58\micron$ and the CO feature at $1.62\micron$ both show tight trends with $\teff$, albeit ones that deviate from those expected from synthetic spectra. 

The EWs we observe deviate most strongly from the EWs we calculate from synthetic spectra for features where the models indicate a strong metallicity dependence in the EWs: \ion{K}{1} at $1.52\micron$, \ion{Si}{1} at $1.58\micron$, CO at $1.61\micron$, and the components of the \ion{Al}{1} doublet at $1.67\micron$. The EWs of our observed spectra match neither the strength of the features in the synthetic spectra, nor the amplitude of their metallicity dependence. This result is similar to what was found by \citet{Rojas-Ayala2012}, who noted disagreement between EWs measured from synthetic and observed spectra for two metallicity-sensitive \K-band features, \ion{Na}{1} at $2.21\micron$ and \ion{Ca}{1} at $2.26\micron$. This suggests that there may be issue with the treatment of metal abundances in M dwarf atmosphere models. 

Lastly, we note that the \K-band spectral indices we measure in our observed spectra are broadly in agreement with those in the synthetic spectra.


\subsection{Empirical calibrations for stellar parameters using spectral features}

We first conducted a principal component analysis (PCA), looking for the strongest correlations between the EW measurements of our calibration sample. We then fit for $\teff$, radius, and luminosity using a linear combination of between one and five principal components. These fits were not better than the simple functions we tried later, and we required line ratios to best fit the full temperature range. The PCA was also hampered by the limited metallicity range of the interferometry stars, but may be a more useful tool in the future.

\begin{figure*}
\includegraphics[width=0.5\linewidth]{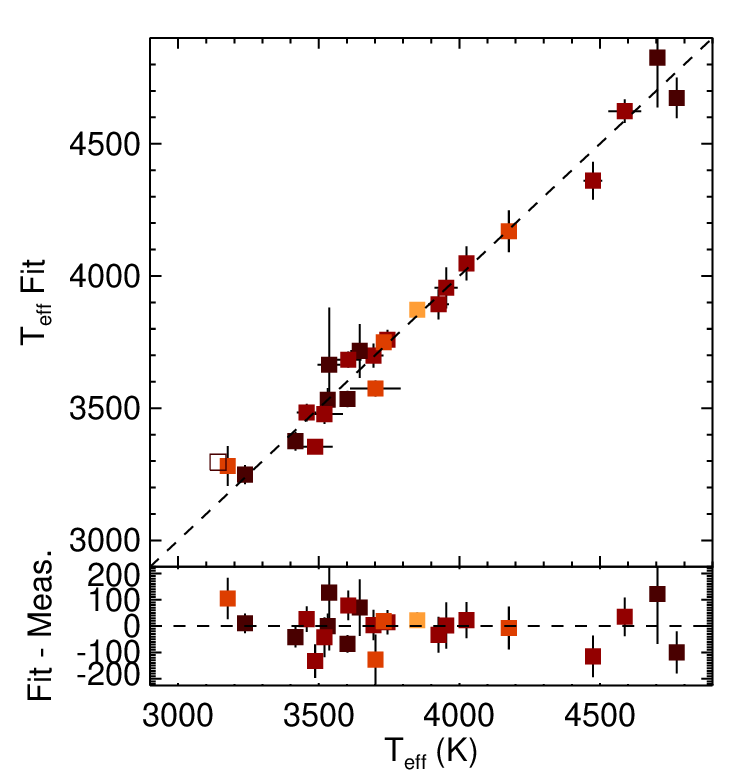}
\includegraphics[width=0.5\linewidth]{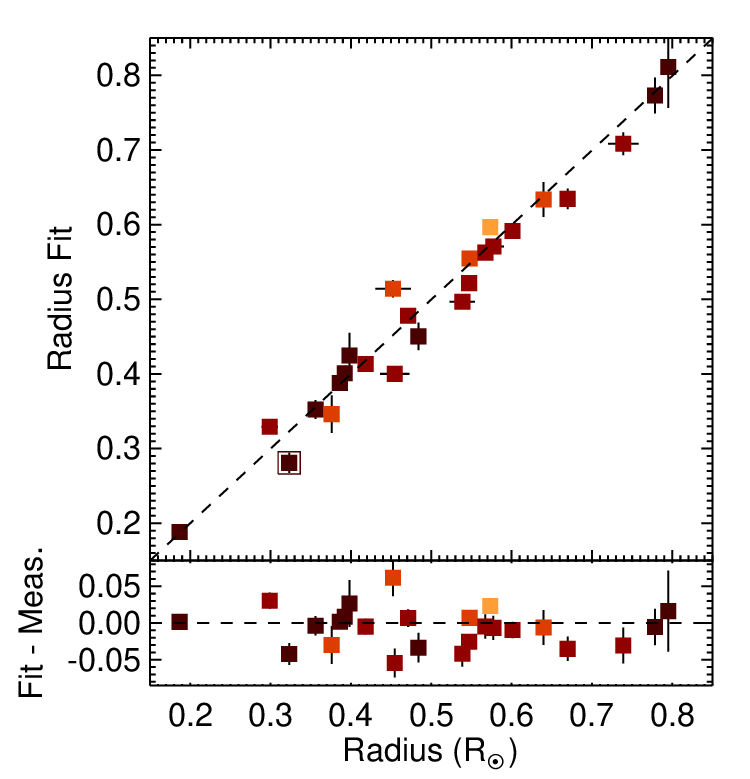}
\includegraphics[width=0.5\linewidth]{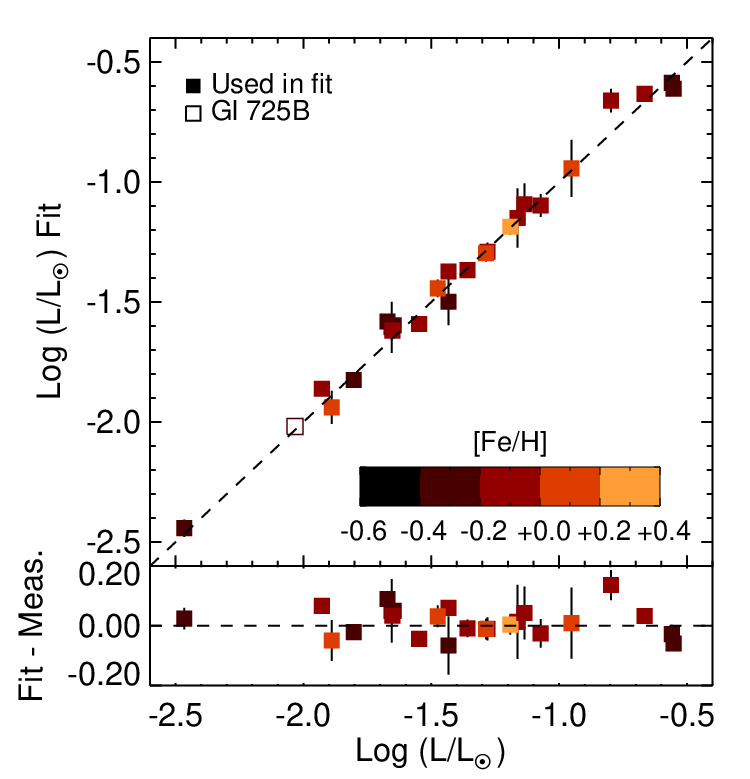}
\caption{Our best-fitting relationships and the residuals for $\teff$ (top left), radius (top right), and $\lbol$ (bottom left). The horizontal axes shows the directly-measured stellar parameter, using the updated values from \citet{Mann2013a}. In the top plot of each panel, the vertical axis shows the stellar parameter we infer from our best fits; in the bottom plot, the vertical axis shows the residuals between our best-fitting values and the directly-measured values. The $\feh$ is indicated by the color of each data point. Gl 725B - used only in our radius fit - is indicated by an open square.
\label{Fig:best-fits}}
\end{figure*}

We investigate simple parameterizations of two or three EWs or EW ratios and simple functions of one EW. The multi-line functions we test (using $x$, $y$, and $z$ to represent an EW measurement or ratio of measurements) are $ax + by$, $ax + by + cz$ and $ax + bx^2 + cy$. The single-line functions we test are $ax + b/x$, $ax+b\sqrt{x}$, and $ax+bx^2$. We restrict our fits to a single NIR band, and perform a comprehensive search across the possible combinations of features. We use least-squares regression to determine the best-fitting parameters for each combination and the Bayesian Information Criterion (BIC) to quantitatively compare each fit. 
After removing fits that qualitatively show systematics in the residuals, we select the fit with the lowest BIC. Our best fits are shown in Figure \ref{Fig:best-fits}.

We found no viable \K-band relationship and the only viable \J-band relationship has significantly higher scatter than our adopted \H-band relation. Combining features from different regions of the spectrum also did not result in significant improvement. We excluded the \ion{Si}{1}, \ion{K}{1}, and CO ($1.62\micron$) features from our fits because the behavior of these features as observed in our larger samples of MEarth and \kepler\ M dwarfs is not well-represented by the behavior of these features in the interferometric sample, which may be due to metallicity. While empirically these lines can be used to fit stellar parameters as well as those we do use, such fits would have more limited applicability.

In the following equations, we use the element responsible for the spectral feature to indicate an EW measurement, for example, $\mathrm{Al\mbox{-}a (1.67\micron)}$ represents the EW measurement of the blue component of the Al doublet at $1.67\micron$. All EWs are in \AA ngstroms. After each equation, we indicate the standard deviation of the residuals ($\sigma$) and the mean absolute deviation (MAD). Our best fitting relationships are:
\begin{align}
\label{Eq:teff}
\teff/\mathrm{K} 
=& +271.4 \times \mathrm{Al\mbox{-}a (1.67\micron)} \\
& +392.7 \times \mathrm{Mg (1.50\micron)}/ \mathrm{Al\mbox{-}b (1.67\micron)} \nonumber \\ 
& + 2427 \nonumber \\
\sigma_{\teff/\mathrm{K}} = & 73 \nonumber \\ 
\mathrm{MAD}/\mathrm{K} = & 55 \nonumber 
\end{align}

\begin{align}
\label{Eq:rad}
R/\rsun 
=& -0.0489 \times \mathrm{Mg (1.57\micron)} \\
&+0.275 \times \mathrm{Al\mbox{-}a (1.67\micron)} \nonumber \\
&+0.201\times \mathrm{Mg (1.57\micron)}/\mathrm{Al\mbox{-}a (1.67\micron)} \nonumber \\
& -0.216 \nonumber \\
\sigma_{R/\rsun} =& 0.027 \nonumber \\
\mathrm{MAD}/\rsun =& 0.020 \nonumber
\end{align}

\begin{align}
\label{Eq:lum}
\log L/\lsun  
=& +0.832 \times\mathrm{Mg (1.71\micron)} \\
& -0.176 \times \mathrm{[Mg (1.71\micron)]}^2 \nonumber \\
& +0.266 \times \mathrm{Mg (1.50\micron)} \nonumber \\
& -3.491 \nonumber \\
\sigma_{\log{L/\lsun}} =& 0.049 \nonumber \\
\mathrm{MAD}/\mathrm{dex} =& 0.039 \nonumber
\end{align}

We show our best fits in Figure \ref{Fig:best-fits}. We present the stellar parameters we estimate for the interferometric calibration sample in Table \ref{Tab:if}; our EW measurements are included in Table \ref{Tab:ifew}. These fits are calibrated from our sample of 24 calibrators (25 for our radius fit, for which Gl 725B is included, see \S\ref{Sec:intstars}). They are valid for stars with $3200<\teff<4800$K, $0.18<R<0.8\rsun$, and $-2.5<\log L/\lsun<-0.5$. 

We note that these calibrations depend on some of the same spectral features. More fundamentally, spectral features are determined by a star's $\teff$, $\logg$, and $\feh$ -- not by its radius and $\lbol$. It could be argued that it is only appropriate to derive $\teff$, $\logg$, and $\feh$  from spectral features, which could then be used to determine radius or $\lbol$. We opt to present relations calibrated directly to radius and $\lbol$ for several reasons: we do not have masses for these stars or other constraints on $\logg$, efforts are currently underway that will revise the temperatures and luminosities of our calibration sample, and disagreement continues over $\teff$ scales \citep[e.g.][]{Casagrande2014}.

\subsection{Systematic uncertainties in the calibrations}

To asses systematic uncertainties in our calibrations, we performed two bootstrap analyses with 1000 samples each. In each case, we randomly created a new realization of our calibration sample and re-fit our relations using each new realization.  In the first bootstrap analysis, we randomly drew stars from our calibration sample, allowing calibrators to be sampled repeatedly. Despite the limited number of calibrators, our fits are well-constrained at the limits of our calibration.
In the second bootstrap analysis, we randomly permuted the residuals from our best fit, assigning each residual to a new data point. This probes the effect of deviations from our best fit that are not captured in the errors on the measured stellar parameters. We find no systematic deviation in $\teff$. For radius and $\lbol$, small systematic errors are evident in the difference between the stellar parameters we infer from our bootstrapped fits and those we infer from our fiducial fit. For radius, the difference is $0.006\rsun$. For luminosity, the difference is $-0.014\dex$ for the coolest stars ($\log L/\lsun=-2.5$dex) and reaches $-0.021$dex for the hottest ($\log L/\lsun>-1.3$dex). We adopt systematic uncertainties of $0.006\rsun$ for radius and $0.02$dex for $\log L/\lsun$.

We additionally demonstrate the robustness of our calibrations for small stars by removing Gl 699 (Barnard's star), which anchors the radius and luminosity fits, and refit our relations. The effect on the parameters we infer for other stars in negligible, and the relations also provide a good fit if we extrapolate them to Gl 699. Extrapolation gives a radius of $0.21\rsun$ (the measured value is $0.19\rsun$) and a log luminosity of $-2.36$dex (measured: $-2.47$ dex) for Gl 699.

$\teff$, radius and $\lbol$ are not independent parameters and our calibrations should produce consistent values. Looking at the difference between $\log L/\lsun$ and $2\log R/\rsun + 4\log T/5777\mathrm{K}$, the scatter is $0.1$ dex with a systematic offset of $-0.06$ dex. This corresponds to a $7\%$ difference in radius -- if the entire systematic offset is assigned to the radius calibration -- or a $3.5\%$ difference in $\teff$ -- if it is assigned to the temperature calibration.

\begin{deluxetable*}{l r r r  l r r}  
\tablecaption{\label{Tab:if}Inferred parameters for the interferometric sample} \tabletypesize{\footnotesize} 
\setlength{\tabcolsep}{0.08in}
\tablecolumns{7}
\tablehead{\colhead{Star} & \multicolumn{3}{c}{Measured parameters} & \multicolumn{3}{c}{Inferred parameters} \\
& \colhead{$\teff$\tablenotemark{a}} 
& \colhead{Radius\tablenotemark{b}} 
& \colhead{$\lbol$\tablenotemark{a}} 
& \colhead{$\teff$\tablenotemark{c}} 
& \colhead{Radius\tablenotemark{d}} 
& \colhead{$\lbol$\tablenotemark{e}} \\
& \colhead{(K)}
& \colhead{($\rsun$)}
& \colhead{($\log L/\lsun$)}
& \colhead{(K)}
& \colhead{($\rsun$)}
& \colhead{($\log L/\lsun$)}
}
\startdata
Gl 725B\tablenotemark{f}&$3142\pm 29$&$0.3232\pm0.0061$&$-2.0329\pm 0.0052$&($3295\pm 81$)&$0.281\pm0.030$&($-2.02\pm 0.06$)\\
Gl 876\tablenotemark{g}&$3176\pm 20$&$0.3761\pm0.0059$&$-1.8893\pm 0.0027$&$3281\pm105$&$0.346\pm0.037$&$-1.94\pm 0.08$\\
Gl 699&$3238\pm 11$&$0.1869\pm0.0012$&$-2.4660\pm 0.0038$&$3248\pm 81$&$0.188\pm0.029$&$-2.44\pm 0.06$\\
Gl 725A&$3417\pm 17$&$0.3561\pm0.0039$&$-1.8033\pm 0.0052$&$3375\pm 81$&$0.352\pm0.030$&$-1.82\pm 0.05$\\
Gl 687&$3457\pm 35$&$0.4183\pm0.0070$&$-1.6521\pm 0.0086$&$3483\pm 80$&$0.413\pm0.028$&$-1.62\pm 0.05$\\
Gl 581&$3487\pm 62$&$0.2990\pm0.0100$&$-1.9278\pm 0.0077$&$3354\pm 74$&$0.329\pm0.027$&$-1.86\pm 0.05$\\
Gl 436&$3520\pm 66$&$0.4546\pm0.0182$&$-1.5476\pm 0.0110$&$3477\pm 81$&$0.400\pm0.028$&$-1.59\pm 0.05$\\
Gl 411&$3532\pm 17$&$0.3924\pm0.0033$&$-1.6708\pm 0.0061$&$3532\pm 85$&$0.401\pm0.029$&$-1.58\pm 0.05$\\
Gl 412A&$3537\pm 41$&$0.3982\pm0.0091$&$-1.6548\pm 0.0047$&$3664\pm227$&$0.425\pm0.041$&$-1.61\pm 0.12$\\
Gl 15A&$3602\pm 13$&$0.3863\pm0.0021$&$-1.6467\pm 0.0052$&$3534\pm 79$&$0.388\pm0.028$&$-1.60\pm 0.05$\\
Gl 649\tablenotemark{g}&$3604\pm 46$&$0.5387\pm0.0157$&$-1.3586\pm 0.0023$&$3683\pm 79$&$0.497\pm0.028$&$-1.37\pm 0.06$\\
Gl 526&$3646\pm 34$&$0.4840\pm0.0084$&$-1.4325\pm 0.0060$&$3716\pm125$&$0.450\pm0.033$&$-1.50\pm 0.11$\\
Gl 887&$3695\pm 35$&$0.4712\pm0.0086$&$-1.4325\pm 0.0088$&$3698\pm 86$&$0.478\pm0.028$&$-1.37\pm 0.05$\\
Gl 176\tablenotemark{g}&$3701\pm 90$&$0.4525\pm0.0221$&$-1.4746\pm 0.0034$&$3574\pm 78$&$0.514\pm0.029$&$-1.44\pm 0.06$\\
Gl 880&$3731\pm 16$&$0.5477\pm0.0048$&$-1.2856\pm 0.0049$&$3749\pm 76$&$0.555\pm0.028$&$-1.30\pm 0.06$\\
Gl 809&$3744\pm 27$&$0.5472\pm0.0067$&$-1.2798\pm 0.0057$&$3758\pm 82$&$0.522\pm0.028$&$-1.29\pm 0.06$\\
Gl 205&$3850\pm 22$&$0.5735\pm0.0044$&$-1.1905\pm 0.0094$&$3872\pm 75$&$0.597\pm0.027$&$-1.19\pm 0.06$\\
Gl 338B&$3926\pm 37$&$0.5673\pm0.0137$&$-1.1627\pm 0.0145$&$3892\pm 92$&$0.562\pm0.028$&$-1.15\pm 0.13$\\
Gl 338A&$3953\pm 41$&$0.5773\pm0.0131$&$-1.1357\pm 0.0164$&$3955\pm106$&$0.571\pm0.029$&$-1.09\pm 0.10$\\
Gl 820B&$4025\pm 24$&$0.6010\pm0.0072$&$-1.0722\pm 0.0064$&$4047\pm 97$&$0.591\pm0.028$&$-1.10\pm 0.07$\\
Gl 380&$4176\pm 19$&$0.6398\pm0.0046$&$-0.9518\pm 0.0065$&$4168\pm107$&$0.634\pm0.036$&$-0.94\pm 0.13$\\
Gl 702B&$4475\pm 33$&$0.6697\pm0.0089$&$-0.7972\pm 0.0108$&$4360\pm102$&$0.635\pm0.030$&$-0.66\pm 0.07$\\
Gl 570A&$4588\pm 58$&$0.7390\pm0.0190$&$-0.6654\pm 0.0064$&$4623\pm 85$&$0.708\pm0.031$&$-0.63\pm 0.05$\\
Gl 105A&$4704\pm 21$&$0.7949\pm0.0062$&$-0.5589\pm 0.0056$&$4825\pm202$&$0.811\pm0.061$&$-0.59\pm 0.06$\\
Gl 892&$4773\pm 20$&$0.7784\pm0.0053$&$-0.5521\pm 0.0060$&$4673\pm106$&$0.773\pm0.036$&$-0.61\pm 0.05$\\
\enddata
\tablenotetext{a}{Calculated from the interferometric radius and bolometric flux as describe in \S\ref{Sec:intstars}. 
Values are from \citet{Mann2013a} Table 1 unless otherwise noted.}
\tablenotetext{b}{Measured from interferometry; 
see \citet{Boyajian2012} Table 6 for references unless otherwise noted.}
\tablenotetext{c}{Inferred from EWs using Equation \ref{Eq:teff}; fit shown in Figure \ref{Fig:best-fits}, top left.}
\tablenotetext{d}{Inferred from EWs using Equation \ref{Eq:rad}; fit shown in Figure \ref{Fig:best-fits}, top right.}
\tablenotetext{e}{Inferred from EWs using Equation \ref{Eq:lum}; fit shown in Figure \ref{Fig:best-fits}, bottom left.}
\tablenotetext{f}{Gl 725B was excluded from our $\teff$ and $\lbol$ calibrations, as discussed in \S\ref{Sec:intstars}. The $\teff$ and $\lbol$ we infer from Equations \ref{Eq:teff} and \ref{Eq:lum} for this star are nevertheless included in this table.}
\tablenotetext{g}{Values for interferometric radii are from \citet{vonBraun2014}; updated $\teff$ and $\lbol$ from this work.}

\end{deluxetable*}

\begin{deluxetable}{l}  
\tablecaption{\label{Tab:ifew}Measured quantities for the interferometric sample} \tabletypesize{\tiny} 
\startdata
This table is available in the online version of this article.
\enddata
\end{deluxetable}

\section{Application to M dwarfs from MEarth}\label{Sec:mearth}

\begin{deluxetable}{l}
\setlength{\tabcolsep}{0.03in}
\tablecaption{\label{Tab:allstars}Measured quantities and inferred stellar parameters for M dwarfs in the MEarth sample} \tabletypesize{\tiny} 
\tablecolumns{1}
\startdata
This table is available in the online version of this article.
\enddata
\end{deluxetable}

We applied our calibrations for $\teff$, radius and luminosity to the sample of MEarth M dwarfs observed by \cite{Newton2014}.

We present measurements of the EWs used in our best fits, the $K$-band temperature index from \citet{Mann2013a}, and our inferred stellar parameters in Table \ref{Tab:allstars}. The quoted errors on our stellar parameters are random errors (propagated from EW errors) added in quadrature to the standard deviation of the residuals in our best fits. The errors do not include our adopted systematic uncertainties, which are $0.006\rsun$ for radius, and $0.02$dex for $\log L/\lsun$. We found no evidence of systematic uncertainties for $\teff$.

The latest spectral type represented in our interferometry calibration sample is M4V (Gl 699, Barnard's star). The MEarth targets predominantly have temperatures and radii at the extreme end of the calibration range, while close to half of sample have EWs that indicate that they are cooler than Gl 699. While our calibrations are not valid for these cool stars, they remain well-behaved for late M dwarfs and are useful diagnostics of stellar properties. We therefore report estimated stellar properties for stars beyond the limits of our calibration, but caution that these values may only be used assess the properties of stars relative to one another.

We limit application of our temperature calibration to objects with small uncertainties in the EWs: when EW uncertainties are a large fraction of the measurement, their ratios have asymmetric error distributions, and we find that the temperatures of cool stars with large errors are systematically hotter than those of similar stars with small errors. Therefore, we do not report estimated temperatures for stars for which the contribution from EW uncertainties (random error) to the total error exceeds $100$K. We found no evidence for similar effects in our radius and luminosity calibrations.

\subsection{Using luminosities to revisit the metallicities of the MEarth sample}\label{Sec:metallicities}   

\citet{Newton2014} estimated the metallicities of the M dwarfs in our sample from the EW of the \ion{Na}{1} feature at $2.2\micron$ in the \K-band of IRTF spectra \citep[see also][]{Rojas-Ayala2010,Rojas-Ayala2012,Terrien2012}. Their empirical relation was based on M dwarfs with NIR spectral types M5V and earlier. \citet{Mann2013} constructed a similar calibration for K5-M5 dwarfs, using a combination of the Na feature and other NIR spectral features. \citet{Mann2014} used early-late M dwarf pairs to bootstrap a calibration valid for M dwarfs with spectral types M4V-M9V and showed that \citet{Newton2014} and other previous works either over- or underestimated the metallicities of M7 to M9 dwarfs. In the following, we make use of our estimated stellar luminosities to investigate the best metallicity calibration to use on our sample of stars.

\begin{figure}
\includegraphics[width=\linewidth]{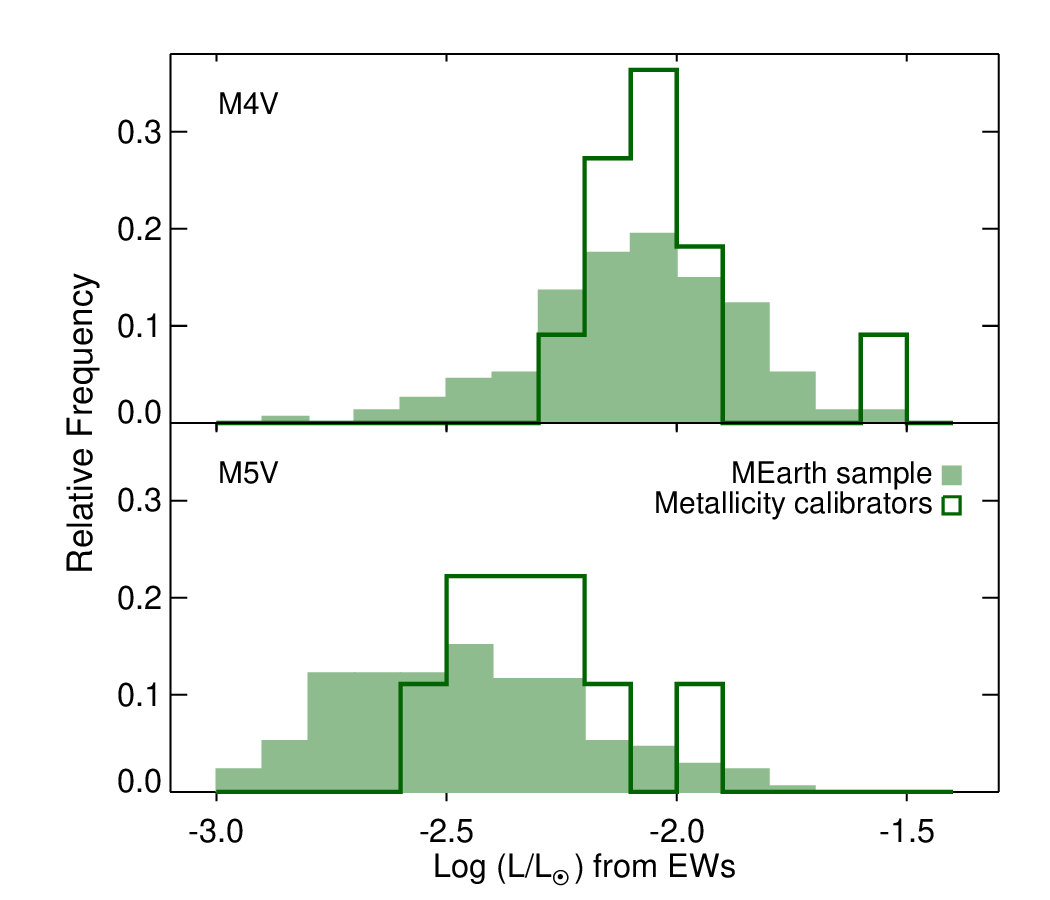}
\caption{The distribution of inferred luminosities for M4 dwarfs (top panel) and for M5 dwarfs (bottom panel). M dwarfs in the MEarth sample are represented by the light green shaded histogram. The M dwarfs in the \citet{Newton2014} metallicity calibration sample are represented by the dark green histogram.
\label{Fig:lum-hist}}
\end{figure}

\citet{Newton2014} looked for systematic trends in the metallicities of stars with spectral type, which they determined by matching each NIR spectrum to that of a spectral standard. They found that the M5-M7 dwarfs in their sample appeared metal-rich by $0.1$ dex relative to earlier stars (Figure 13 in their work). Because M5 dwarfs comprised a quarter of their metallicity calibration sample, they assumed the metallicities of the MEarth M5 dwarfs - and by extension the M6 and M7 dwarfs - were estimated accurately. However, M spectral types are a coarse binning in stellar parameters. In Figure \ref{Fig:lum-hist}, we show that while the M4V metallicity calibrators used by \citet{Newton2014} are typical of the M4 dwarfs in the MEarth sample, $70\%$ of the M5 dwarfs have luminosities lower than the median luminosity of the M5V metallicity calibrators. The calibration from \citet{Newton2014} may not be valid for all M dwarfs assigned an M5V spectral type.

\begin{figure}
\includegraphics[width=\linewidth]{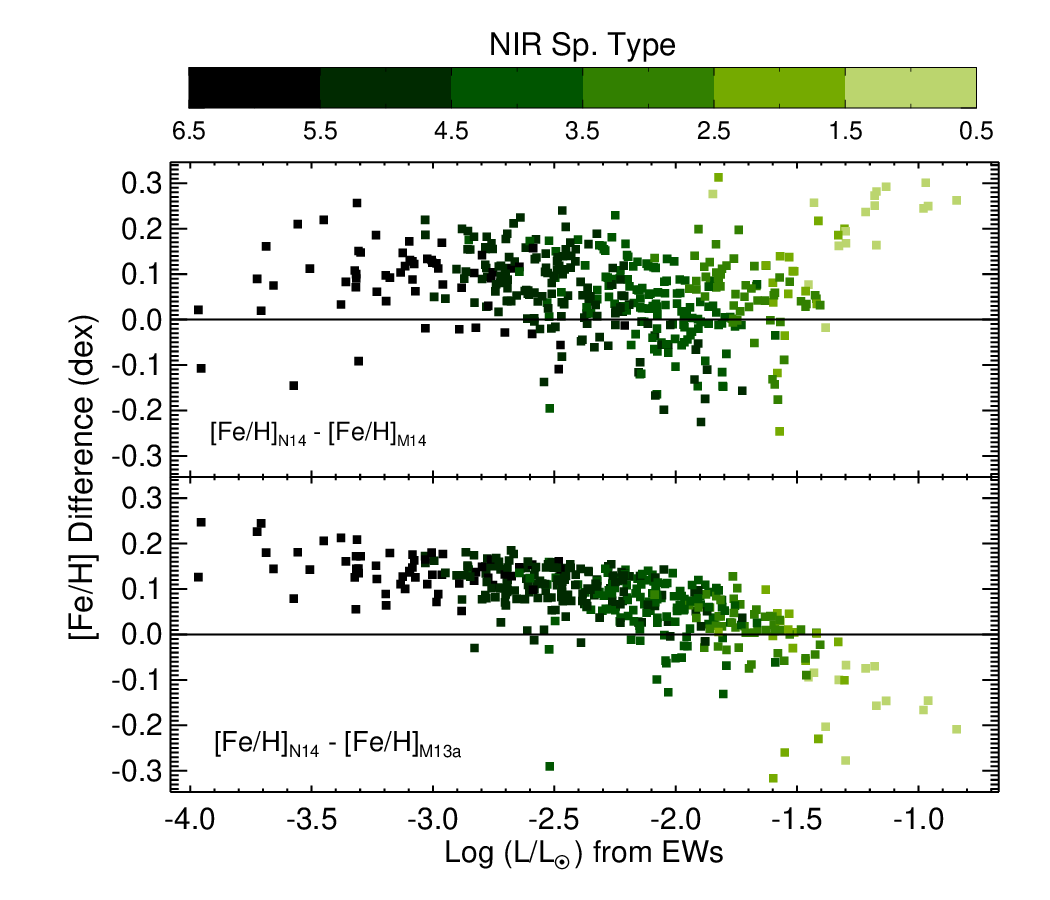}
\caption{The difference between literature $\feh$ values versus stellar luminosity. We use relations from three works: \citet[M13a]{Mann2013}, \citet[N14]{Newton2014}, and \citet[M14]{Mann2014}. The \cite{Mann2013} relation is calibrated for M4V stars and earlier, while the \citet{Mann2014} relation is calibrated for stars M4V and later. Points are colored by the NIR spectral type assigned by eye by \citet{Newton2014}.
\label{Fig:feh-lum}}
\end{figure}

In Figure \ref{Fig:feh-lum}, we compare $\feh$ values measured for the MEarth M dwarfs using the calibrations from \citet{Newton2014}, \citet[][]{Mann2013}, and \citet[][]{Mann2014}. We only show M dwarfs whose metallicities, as estimated from the \citet{Newton2014} calibration, are less than $0.25$dex, because the \cite{Newton2014} calibration saturates for more metal-rich stars. This demonstrates that while the \citet{Newton2014} calibration is generally valid for M2-M4V stars, it over- and under-estimates the metallicities of later and earlier stars, respectively.

We compare the \citet{Newton2014} calibration to the \citet{Mann2014} calibration, which is applicable to late M dwarfs, in the top panel of Figure \ref{Fig:feh-lum}. This demonstrates that \citet{Newton2014} overestimates the metallicities of M dwarfs with $\log L/\lsun<-2.25$, which includes the majority of stars M5V and later. The two calibrations agree for stars with $-2.25<\log L/\lsun<-1.75$ (approximately M4V), with a median offset of $0.03$dex. For hotter stars, the \citet{Mann2014} calibration is not applicable, and the estimated metallicities deviate significantly. In the bottom panel of Figure \ref{Fig:feh-lum}, we find an almost linear trend with luminosity when we compare the \citet{Newton2014} to the \citet{Mann2013} calibration, which is applicable to early M dwarfs. The two calibrations agree most closely for $-2.0 < \log L/\lsun < -1.5$, (approximately M2V-M3V). For earlier stars, the \citet{Newton2014} calibration underestimates metallicities, a finding anticipated in that work.

We update the metallicities of the MEarth M dwarfs by stitching together the \citet{Mann2013} and \citet{Mann2014} relations. For stars with $\log L/\lsun<-1.75$, we use the \citet{Mann2014} relation, which was calibrated specifically for late-type dwarfs. For stars with $\log L/\lsun>-1.75$, we use the \citet{Mann2013} relation, for which the calibration was dominated by early M dwarfs. We use the code provided by A. Mann\footnote{https://github.com/awmann/metal} to calculate metallicities. We note that the method we use to measure EWs (over-sampling and using the trapezoidal rule to numerically integrate the flux beneath the pseudo-continuum), while more robust than the IDL routines \texttt{TOTAL} or \texttt{SUM} without over-sampling, produces different (typically larger) EW measurements.


\subsection{Comparison to the Mann et al.~temperature and radius calibrations}\label{Sec:mann}

\citet{Mann2013a} developed empirical relations for cool dwarf temperatures that are based on spectral indices in visual and infrared bands. As we have done in this work, they used stars with interferometric measurements to calibrate the relationships. The temperature-sensitive indices are ratios between the median flux in three wavelength windows (Equation 13 in their paper) and quantify the curvature of the spectrum in each band. \citet{Mann2013a} chose the continuum windows to minimize the scatter in the resulting temperature calibration. 

\begin{figure}
\includegraphics[width=\linewidth]{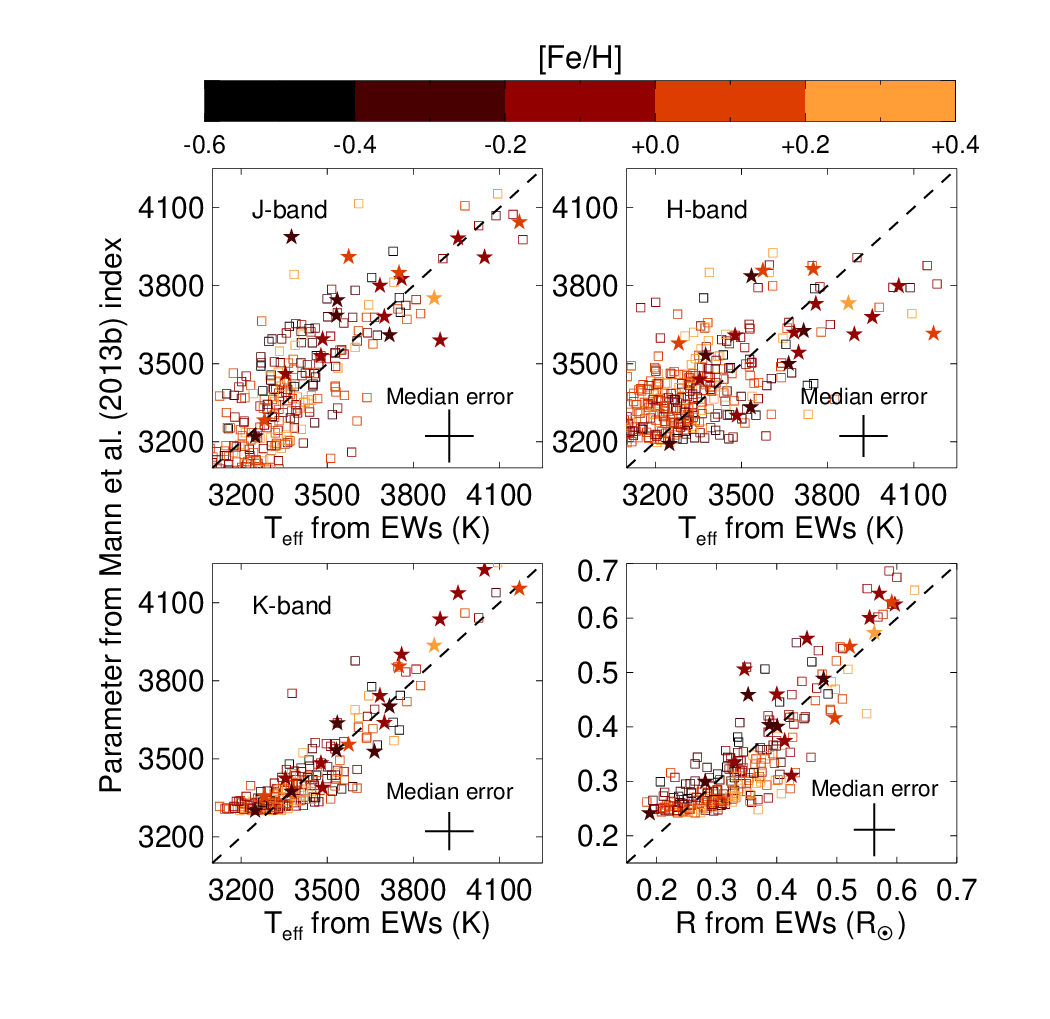}
\caption{A comparison of the calibrations from \citet{Mann2013a} and from this work, showing the $\teff$ and radii estimated for the M dwarfs from \citet{Newton2014}. On the horizontal axis, we show the parameters we infer using the EW-based methods we develop in this work. On the vertical axis, we show the parameters we infer using the index-based methods from \citet{Mann2013a}. We use the polynomial presented in that work to convert temperatures estimated from the \K-band index to radii. The color of the points indicate the stars' metallicities.
\label{Fig:mann-compare}}
\end{figure}

We compare the stellar parameters we estimated for the MEarth M dwarfs to those we calculated using the \citet{Mann2013a} \J-, \H-, and \K-band $\teff$ calibrations in Figure \ref{Fig:mann-compare}. We then applied their temperature-radius polynomial (Equation 6 in their work) to convert the \K-band temperatures to radii. Accurate transformation requires additional significant figures, included here for completeness:
\begin{align}
R/\rsun =& -16.883175 +1.1835396\times10^{-2} \times (\teff/K) \nonumber \\
&-2.7087196\times10^{-6}\times(\teff/K)^2 \\
&+ 2.1050281\times10^{-10}\times(\teff/K)^3 \nonumber
\end{align}

We limit our comparison to those stars within the limits of the calibrations in question; only these objects are included in Figure \ref{Fig:mann-compare}. We note that the \citet{Mann2013a} \K-band calibration saturates for stars with $\teff<3300$K, so there is a minimum value for the temperatures derived from this relation.

The \citet{Mann2013a} \J- and \H-band indices are not good predictors of $\teff$. The differences between the $\teff$ inferred from our two calibrations have standard deviations ($\sigma_{\Delta T}$) larger than expected from the errors in the calibrations: $\sigma_{\Delta T}=140$K for the \J-band calibration and $170$K for the \H-band calibration. This is because the wavelength windows for these indices fall in regions with strong telluric absorption. Temperatures from the \K-band index and our EW-based calibration agree within the errors of the calibrations, with $\sigma_{\Delta T}=90$K, although for $3500<\teff<3800$, the median temperature we infer from EWs is $40$K hotter.

The radii we infer directly from EWs and the radii calculated from the \K-band using the temperature-radius relation have $\sigma_{\Delta R}=0.05\rsun$. However, there is evidence for a systematic offset for stars with $0.3<R/\rsun<0.4$, where the median offset is $0.045\rsun$ (12\%), with radii calculated from EWs being larger. 
We believe this is due to systematics in the \citet{Mann2013a} relations: the \K-band temperature relation predicts temperatures that are too cool by $50-100$K for the three interferometry stars with temperatures around $3500$K (Figure 11 in their work). Because the slope of the temperature-radius relation is steep for stars of this size, temperatures that are $50$-$100$K too cool result in radii that are $0.03$-$0.06\rsun$ (15\%) too small. This is consistent with the differences we find and supports the temperatures and radii we infer from EWs. Additionally, the temperature-radius relation has larger scatter for stars of this size relative to hotter/larger stars (Figure 4 in their work). 

Intriguingly, we find that stars with $\feh<-0.2$dex are assigned larger radii by the \citet{Mann2013a} relations than stars are with $\feh>+0.2$dex. The mean difference is $0.05\rsun$. The effect persists when applying the temperature-radius relation to temperatures inferred using our EW relations, so the difference must either be a result of the \citet{Mann2013a} temperature-radius relation or of our radius calibration. We suggest that the root cause is the temperature-radius relation for two reasons: (1) we expect a metallicity dependence in the temperature-radius relation from theory (one has yet to present itself in observations of the interferometry stars, but the metallicity range spanned by these data is narrow) and (2) we do not see a metallicity dependence when we consider inferred radius as a function of $\mk$ (\S\ref{Sec:mk}).

\subsection{Trends between absolute $\mk$ and inferred stellar parameters}\label{Sec:mk}

\begin{figure*}
\includegraphics[width=0.5\linewidth]{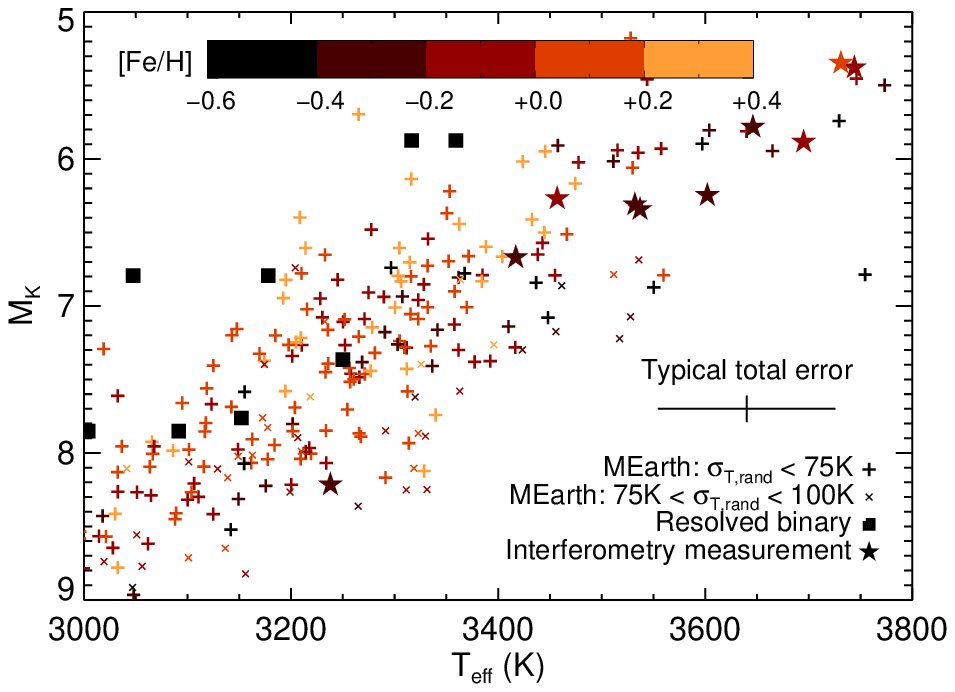}
\includegraphics[width=0.5\linewidth]{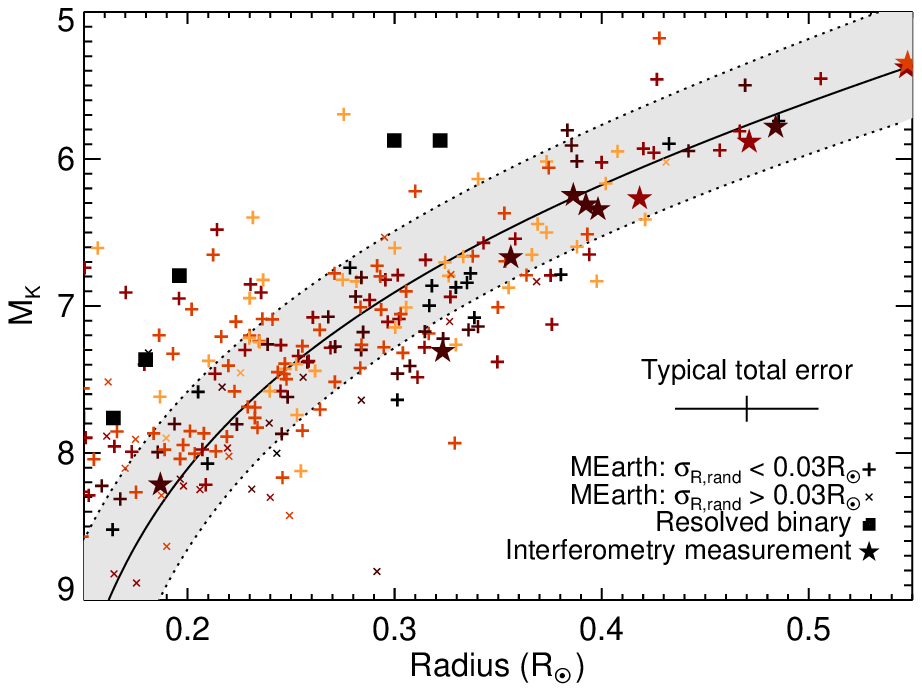}
\includegraphics[width=0.5\linewidth]{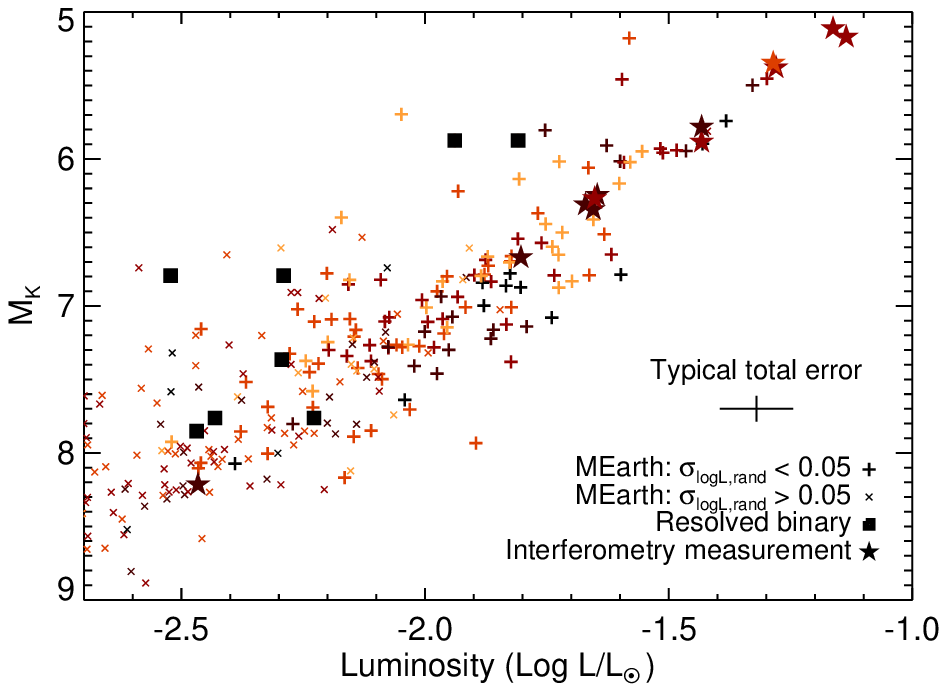}
\caption{Absolute \K magnitude versus inferred $\teff$, radius, or $\lbol$ for MEarth M dwarfs from \citet{Newton2014}. Included are stars with $\sigma_\mathrm{MK}<0.2$ mag; the symbol indicates the size of the error on stellar parameter (plus symbols for smaller errors, crosses for larger errors).
We over-plot the properties of the interferometry stars (filled stars); we show the measured radii and the temperatures and luminosities calculated in \citet{Mann2013a} and in this work. The $\feh$ of each data point is indicated by its color. We also indicate binaries that are unresolved in 2MASS but resolved in our IRTF observations (filled black squares). In the radius plot (upper right), we also include an $\mk$-radius relation, calculated from the \citet{DelfosseX.2000} $\mk$-mass relation and the \citet{Boyajian2012} mass-radius relation. The shaded region indicates $\pm15\%$.
\label{Fig:parammag}}
\end{figure*}

We consider the relationship between $\mk$ and the $\teff$, radius, or $\lbol$ estimated from EWs in Figure~\ref{Fig:parammag}. NIR photometry is from 2MASS \citep{Skrutskie2006}. We exclude $53$ objects with $\sigma_{\mk}>0.2$ and $9$ objects lacking high quality magnitudes from 2MASS (for which the \texttt{qual\_flag} is anything other than AAA). For $\teff$, we also exclude those objects that have random errors on $\teff$ $>100$K as discussed at the beginning of \S\ref{Sec:mearth}. We fit a quadratic for $\mk$ as a function of stellar temperature, radius, or log luminosity; the standard deviation in $\mk$ is about $0.5$ mag for each fit. Because $\mk$ is an independent indicator of stellar parameters -- for example, it was tied to stellar mass by \citet{DelfosseX.2000} -- the existence of a clear relationship between $\mk$ and the $\teff$, radii, and $\lbol$ we infer provides additional validation of our method. 

Visually, our plot of $\mk$ versus $\teff$ shows the largest scatter. Large variations in $\teff$ and $\mk$ are also seen in the Dartmouth models \citep[]{Dotter2008a, Feiden2011} which predict that a star with $\feh=-0.5$ dex and $\teff=3400$ will be $1$ magnitude fainter than a star with the same temperature but with $\feh=+0.3$ dex.  While neither our $\teff$ estimates nor the interferometric measurements indicate such a metallicity dependence, the models still demonstrate the strong influence atmospheric opacities can have on $\teff$ (they predict less sensitivity for $\mk$ versus radius or $\lbol$). Unresolved binaries also contribute significantly to the scatter, which we discuss in \S\ref{Sec:binaries}.

\subsection{Identifying over-luminous objects}\label{Sec:binaries}

An object that is an unresolved multiple will have an $\mk$ smaller than that of a single star with the same value of $\lbol$ predicted from our Equation \ref{Eq:lum}: the EWs of spectral features are largely unchanged by the object's multiplicity (though for very tight binaries features might appear broadened), while the object appears brighter. Our sample contains binaries that are unresolved in 2MASS but for which independent spectra of the components were obtained: when two stars could be identified in the SpeX guider, \citet{Newton2014} aligned the slit along both components and extracted each object separately. As expected, many of these objects are brighter than single stars with similar spectrally-inferred stellar parameters. The diagram of $\mk$ versus $\lbol$ shows the excess scatter at brighter magnitudes most clearly.

To quantify the likelihood of an object being a multiple, we calculate the significance of the magnitude offset ($S$) between the object and a quadratic that describes our empirical $\mk$-luminosity sequence. We fit $\mk$ as a function of $\log L/\lsun$ for stars with $\log L/\lsun>-2.5$, which is the limit of our calibration, using $\sigma_{\mk}$ as our measurement error. We then exclude those objects whose magnitudes are brighter than the best-fitting magnitude by more than the $0.52$ mag, which is the standard deviation in the residuals, and re-fit the data. We extrapolate the fit to stars with fainter luminosities. $S$ is then given by:
\begin{align}\label{Eq:sig}
S = \frac{ \mk - \mathrm{M}_\mathrm{K,fit}} {\sqrt{\sigma_{\mk}^2 + \sigma_{\mathrm{M}_\mathrm{K,fit}}^2}}
\end{align}

\begin{figure}
\includegraphics[width=\linewidth]{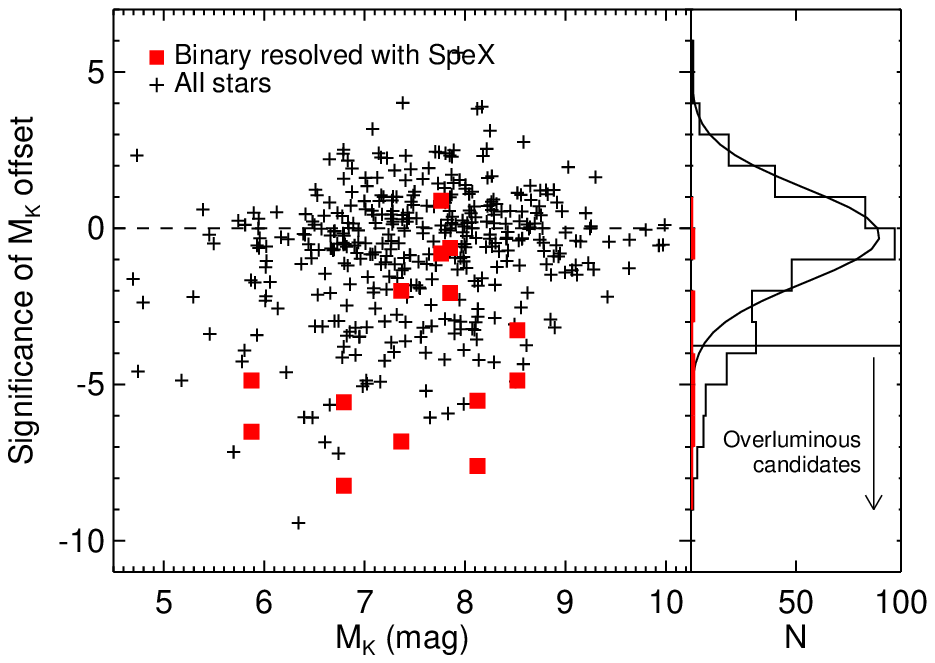}
\caption{Significance $S$ versus absolute K magnitude for the MEarth M dwarfs. $S$ is the offset in magnitudes from our empirical $M_K$-log luminosity main sequence divided by the error, which includes the error in absolute magnitude and in inferred luminosity. On the right side of the figure, we include a histogram of the significances for our entire sample (black histogram) and the best-fitting Gaussian, which has a width of $1.5$ (solid curve). We also show the histogram of significances for the IRTF-resolved binaries (solid red histogram). In the histogram panel, we indicate the $2.5\sigma$ cut we use to select candidate over-luminous objects, which are listed in Table \ref{Tab:binaries}.
\label{Fig:binaries}}
\end{figure}

We show the distribution of $S$-values in Figure \ref{Fig:binaries}, which is well-modeled by a Gaussian with a width of $\sigma=1.5$. An obvious feature is an overabundance of stars that appear too bright for the $\lbol$ we infer (objects with $S<0$), and we identify objects with very negative $S$-values as potential multiples. We note that the width of the Gaussian indicates that our estimates of the errors do not account for all of the scatter in the relation. One possible contributor is binaries with more extreme mass ratios, as the secondary would increase the combined brightness by a lesser amount. 

We select those stars with $S<-3.76$ ($-2.5\sigma$) as candidate multiples. We include the list of candidates in Table \ref{Tab:binaries}, which we note does not include the systems identified as visual binaries by \citet{Newton2014}, which can be found in that work. $18$ of these stars are in fact known binaries that were not resolved in our IRTF observations, as indicated in the table. Others may be previously-unidentified multiples. While youth is another possibility for explaining their brightnesses, a candidate over-luminous object would need to be younger than several hundred Myr to show an enhanced \K-band magnitude, which is unlikely for a field M dwarf.
 
\begin{deluxetable}{l r l l}
\tablecaption{\label{Tab:binaries}Candidate overluminous objects}
\tablecolumns{4}
\tablehead{ \colhead{Object} & \colhead{$\lvert S \rvert$\tablenotemark{a}} &
\colhead{Binary Ref.\tablenotemark{b}} & \colhead{Binary Type\tablenotemark{c}}}
\startdata
LSPM J0008+2050 &4.8&B04& VB \\
LSPM J0028+5022 &4.1&D07& VB \\
LSPM J0105+2829 &4.1&$\ldots$& $\ldots$ \\
LSPM J0111+1526 &4.9&B04& VB \\
LSPM J0159+0331E&4.6&S10& SB2 \\
LSPM J0259+3855 &3.9&L08& VB \\
LSPM J0409+0546 &4.1&L08& VB \\
LSPM J0438+2813 &4.0&B04& VB \\
LSPM J0528+1231 &6.0&$\ldots$&     $\ldots$ \\
LSPM J0540+2448 &5.4&D76& VB \\
LSPM J0711+4329 &4.2&M06& VB \\
LSPM J0736+0704 &5.1&H97& VB \\
LSPM J0810+0109 &3.9&$\ldots$&    $\ldots$ \\
LSPM J0835+1408 &6.0&$\ldots$&    $\ldots$ \\
LSPM J0918+6037W&4.1&$\ldots$&    $\ldots$ \\
LSPM J1000+3155 &4.7&$\ldots$&    $\ldots$ \\
LSPM J1233+0901 &5.7&WDS\tablenotemark{d}& VB \\
LSPM J1331+2916 &7.1&G02& SB2 \\
& & B04& VB \\
LSPM J1332+3059 &4.2&$\ldots$&    $\ldots$ \\
LSPM J1419+0254 &3.7&$\ldots$&    $\ldots$ \\
LSPM J1547+2241 &5.6&$\ldots$&    $\ldots$ \\
LSPM J1555+3512 &4.2&M01& VB \\
LSPM J1604+2331 &5.0&$\ldots$&    $\ldots$ \\
LSPM J1616+5839 &4.8&$\ldots$&    $\ldots$ \\
LSPM J1707+0722 &3.8&P05& VB \\
LSPM J1733+1655 &5.9&$\ldots$&    $\ldots$ \\
LSPM J1841+2447S&6.8&G96& SB2 \\
LSPM J2010+0632 &4.6&S10& SB2  \\
LSPM J2040+1954 &3.9&WDS\tablenotemark{e}& VB  \\
LSPM J2117+6402 &7.1&$\ldots$&  $\ldots$  \\
LSPM J2223+3227 &4.8&W60& VB  \\
\enddata
\tablenotetext{a}{Absolute value of the significance of the magnitude offset from our empirical main sequence as defined by Equation~\ref{Eq:sig}. All objects are over-luminous given their inferred luminosities.}
\tablenotetext{b}{Reference for previous discovery of the object as binary. If no reference is listed, our literature search did not identify that the object is known to be binary. References:~
W60 = \citet{Worley1960};
D76 = \citet{Dahn1976};
G96 = \citet{Gizis1996};
H97 = \citet{Henry1997};
M01 = \citet{McCarthy2001};
G02 = \citet{Gizis2002};
B04 = \citet{Beuzit2004};
P05 = \citet{Pravdo2005};
M06 = \citet{Montagnier2006};
D07 = \citet{Daemgen2007};
L08 = \citet{Law2008};
S10 = \citet{Shkolnik2010}
}
\tablenotetext{c}{Type of binary:~
VB = visual binary;
SB2 = double-lined spectroscopic binary
}
\tablenotetext{d}{This object is the well-known binary Wolf 424 (Gl 473AB), which is identified in the Washington Double Star catalog. We confirmed that its original discovery was by \citet{1938Har}.}
\tablenotetext{e}{This object is listed in the Washington Double Star Catalog as a visual binary, discovered by Riddle~et~al.~(in prep) using Robo-AO, who indicate that it is a chance alignment.}
\end{deluxetable}

\section{Applications to \kepler\ Objects of Interest}\label{Sec:kepler}

We apply our calibrations to the high fidelity sample of 66~KOIs selected as described in Section~\ref{Sec:koiselection}. The KOIs are primarily early M to mid K dwarfs, with a small number of mid M dwarfs. \citet{Muirhead2014} identified three KOIs as visual binaries (two of which meet our sky emission cuts) and KOI 256 as a white dwarf-M dwarf eclipsing binary, which we exclude from this analysis. Unlike for the MEarth objects, the EWs for the KOIs do not approach zero and it is therefore not necessary to use the limits on the errors we adopted for the MEarth objects. However, for ease of comparison, we restrict the discussion in this section to those KOIs with random errors less of than $150$K on $\teff$. This leaves us with $51$ non-visual binaries in our KOI sample.

The cool KOI sample has been the subject of several recent works, and as part of this work, we compare the stellar parameters we infer to those presented in \citet{Dressing2013}, \citet{Mann2013a}, and \citet{Muirhead2014}. 
\citet[]{Dressing2013} matched observed colors to Dartmouth models of different ages, metallicities, and masses. From the best-fitting model, they revised the temperatures and radii for all the M dwarfs in \kepler\ with formal errors typically between $50$ to $100$K for $\teff$ and $0.06\rsun$ for radius.
\citet{Mann2013a} fit flux-calibrated optical spectra to PHOENIX models to determine the $\teff$ of \kepler\ M dwarfs with an estimated error of $57$K, and used relations derived from stars with interferometric measurements to calculate radius and luminosity from $\teff$. Their radius and luminosity errors are $7\%$ and $13\%$, respectively. They refined and tested their method using the interferometric sample, which is also the basis for our work. We note that when fitting spectra without continuum removal, the spectral shape -- and therefore the color -- is an important determinant.
\citet{Muirhead2014} followed the method developed in \citet{Muirhead2012a}, using moderate resolution $K$-band spectra to determine temperatures and metallicities for $103$ cool KOIs. Their temperatures are based on the $\hind$ index \citep{Covey2010}, which was calibrated by \citet[]{Rojas-Ayala2012} as a tracer of $\teff$ using measurements of \btsettl\ model spectra. Typical formal errors for $\teff$ are $65$K.
The authors then inferred the stars' radii by interpolating the $\teff$ and metallicities onto Dartmouth isochrones, achieving a median formal error on stellar radius of $0.06\rsun$. We note that \citet{Muirhead2014} used new stellar models, and provided updated parameters for those objects that were first presented in \citet{Muirhead2012}.

\subsection{Comparison to previous work}

\begin{figure}
\includegraphics[width=\linewidth]{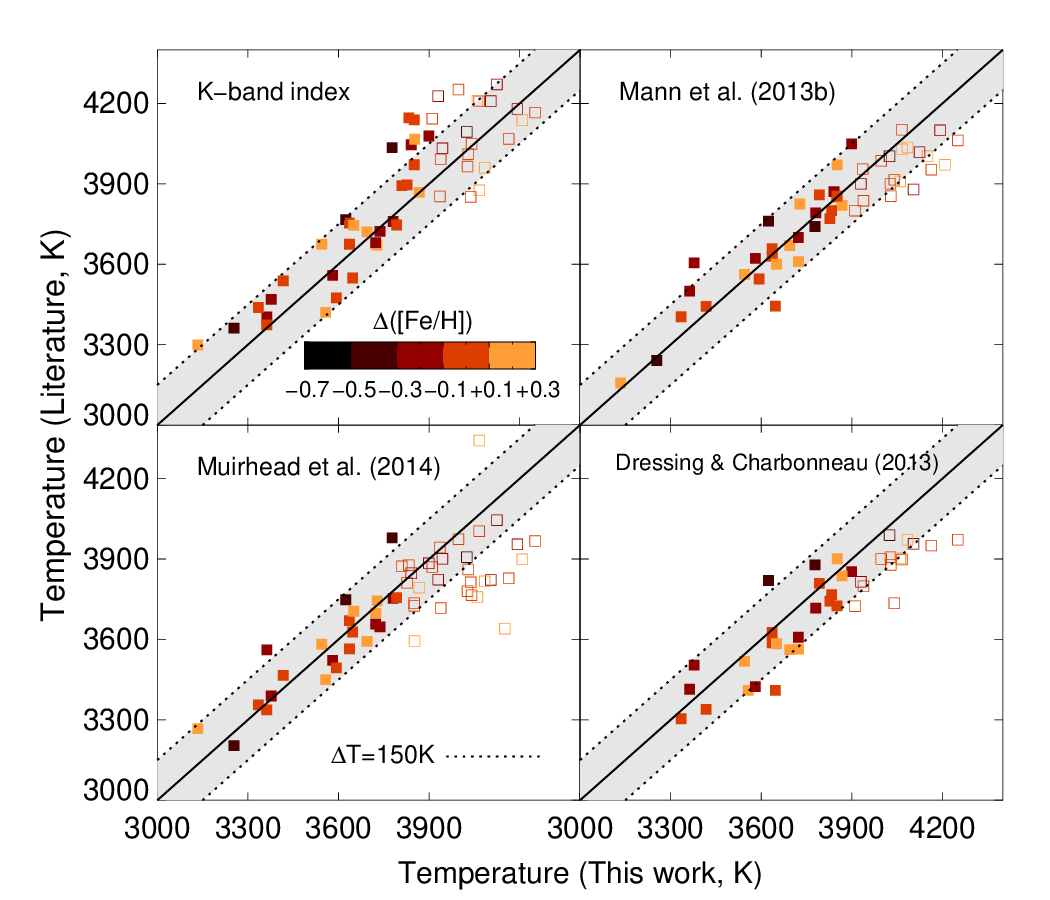}
\caption{Comparison between the temperatures we infer for the KOIs and those inferred from previous works. Our estimates are on the horizontal axis, and those from the literature are on the vertical axis. The literature works are indicated on each panel. We show the one-to-one line and indicate $\pm150$K as the shaded region. Stars with $\teff>3900$K \citep[$\teff>3800$K for][]{Muirhead2014} are shown as open squares; due to selection effects, we limit our discussion to cooler stars.
\label{Fig:koi-temperatures}}
\end{figure}

\begin{figure}
\includegraphics[width=\linewidth]{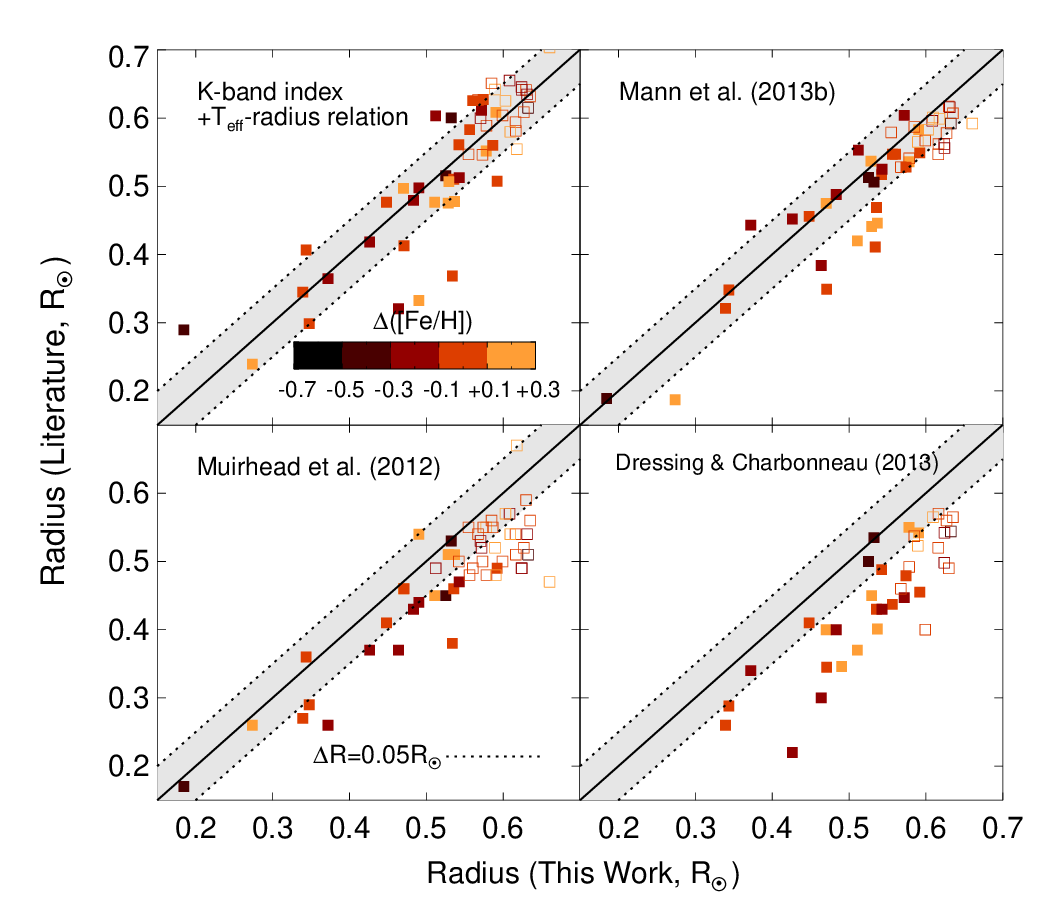}
\caption{Comparison between the radii we infer for the KOIs and those inferred from previous works. Our estimates are on the horizontal axis, and those from the literature on the vertical axis. The literature works are indicated on each panel. We show the one-to-one line and indicate $\pm0.05\rsun$ using the shaded region. Stars with $\teff>3900$K ($>3800$K for \citet{Muirhead2014} are shown as open squares; due to selection effects, we limit our discussion to cooler stars.
\label{Fig:koi-radii}}
\end{figure}

We compare our inferred temperatures and radii to those from previous works in Figures~\ref{Fig:koi-temperatures} and~\ref{Fig:koi-radii}. The literature sources we query are \citet{Mann2013a}, \citet{Muirhead2014}, and \citet{Dressing2013}. We also apply the \K-band calibration from \citet{Mann2013a} to the KOI spectra to estimate $\teff$, after which we use their temperature-radius relation to estimate radii (see \S\ref{Sec:mann}). We adopt the metallicities from \citet{Muirhead2014}.

We cannot make statements about temperature differences above $3900$K. The literature sources we query have their own target selection criteria, each of which places an upper limit of roughly $4000$K on the temperatures of the stars in their sample. This is particularly important for the sample from \citet{Dressing2013}, whose sample is strictly limited to stars with updated temperatures of less than $4000$K. \citet{Mann2013a} and \citet{Muirhead2014} both use color cuts to select red objects, which could also bias the resulting temperatures since their methods are related to spectral colors. Finally, the $\hind$ index used by \citet{Muirhead2014} saturates for $\teff>3800$K.

The temperatures we find generally agree well with those from other works. They are cooler than those inferred from previous spectra-based approaches: the median temperature difference is $80\pm90$K considering the \K-band index, $10\pm60$K considering \citet{Mann2013a}, and $30\pm70$K considering \cite{Muirhead2014}. Compared to \citet{Dressing2013}, the temperatures we infer are hotter by a median of $40\pm90$K. The errors we quote are the median absolute deviations in the temperature differences.

The radii we infer are larger than those found by \citet{Mann2013a}, with a median difference of $0.02\rsun$ and a median absolute deviation of $0.04\rsun$. This difference may be due to their use of a temperature-radius relation, which assumes that all stars at a given temperature have the same radius, but is within reasonable systematic uncertainties for the calibrations.

\begin{figure}
\includegraphics[width=\linewidth]{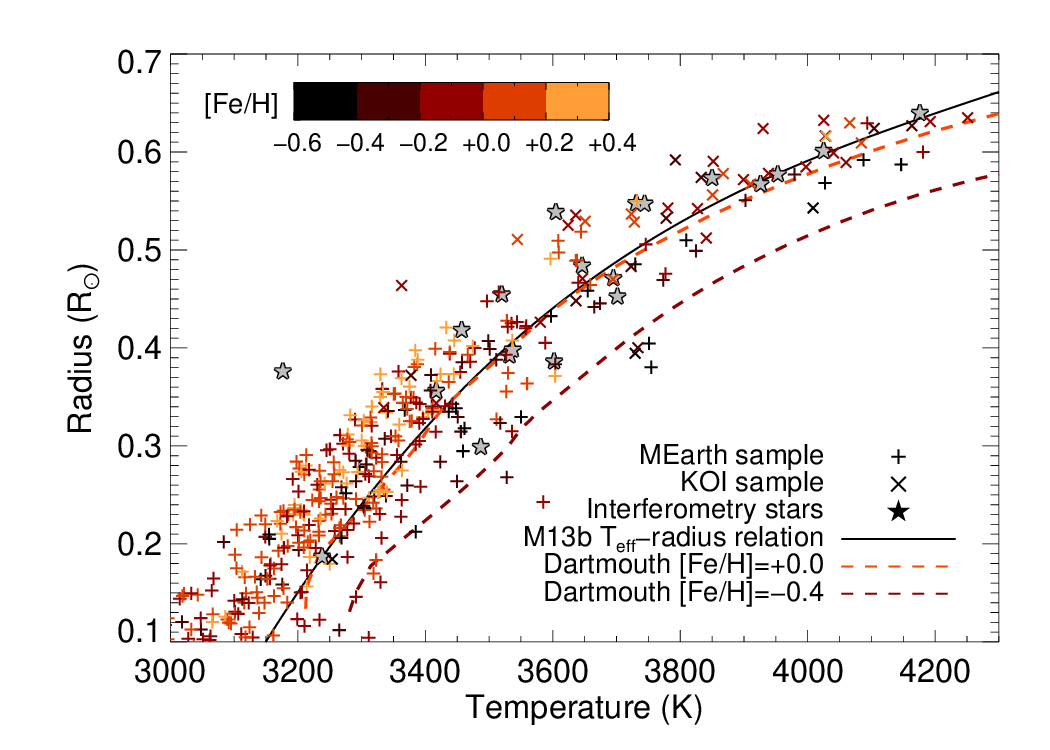}
\caption{Temperature-radius diagram for the stars in this work, showing the measured values for the interferometric sample (gray stars) and the values we infer using EWs for the MEarth sample (plus symbols) and the KOIs (crosses). We color the MEarth and KOI samples by their estimated metallicities. For MEarth, we estimate metallicities using relations from \citet{Mann2013} and \citet{Mann2014}, and for the KOIs we use the metallicities from \citet{Muirhead2014}. We also include the \citet{Mann2013a} temperature-radius relation (solid black line) and 5 Gyr, solar alpha-enhancement Dartmouth isochrones for $\feh=0.0$ (dashed orange line) and $\feh=-0.4$ (dashed dark red line). Gl 725B (measured $\teff=3142$K, $R=0.32\rsun$) was used to calibrate our radius relation but not our temperature relation, and is not shown. The largest outlier from the interferometric sample is Gl 876 (measured $\teff=3176$K, $R=0.38\rsun$).
\label{Fig:teffrad}}
\end{figure}

Our radii are also larger than the radii estimated using methods based on model isochrones, by $0.05\pm0.03\rsun$ relative to \citet{Muirhead2014} and $0.09\pm0.04\rsun$ relative to \citet{Dressing2013}. Because our temperatures are in agreement, the difference in the radii must result from their use of models to estimate radius from temperature. Indeed, \citet[][Figure 14]{Boyajian2012} found that for stars with interferometrically-measured radii between $0.4<R/\rsun<0.6$, the model radii -- predicted by interpolating each stars' $\teff$ onto Dartmouth models -- are smaller than the measured radii by about $10\%$, or $0.05\rsun$. We illustrate the disagreement between observations and models by comparing the temperatures and radii of the interferometric sample to Dartmouth isochrones in Figure \ref{Fig:teffrad}. This effect is sufficient to explain the differences between our radii and those from \citet{Muirhead2014}.

The discrepancy between models and observations also affects the radii from \citet{Dressing2013}. However, the difference between our radii and those from \citet{Dressing2013} is larger because of the metallicities they estimate. In the Dartmouth models, an M dwarf with $\feh=-0.2$ is about $0.04\rsun$ smaller than a solar-metallicity dwarf of the same temperature; such a metallicity dependence is not seen in the temperatures and radii of the interferometry stars. \citet{Dressing2013} estimated sub-solar metallicities for most of the KOIs, whereas the metallicities from \citet{Muirhead2014} are closer to solar (see Figure 7 in Dressing\ \&\ Charbonneau) and \citet{Dressing2013} therefore find smaller radii. This can also be seen in Figure \ref{Fig:teffrad}.
The overall offset between the observed and theoretical temperatures and radii, and the sub-solar metallicities they estimated are what cause our inferred radii to be substantially larger than those reported in \citet{Dressing2013}.


\subsection{Updated stellar and planetary parameters}

We use the data available on the NASA Exoplanet Archive\footnote{http://exoplanetarchive.ipac.caltech.edu/, accessed 2014/06/10} to update the properties of the planet candidates orbiting the cool KOIs. We compare the planets properties that we get using our stellar parameters to those that one would infer using the stellar parameters in the catalog from \citet{Huber2014}. \citet{Huber2014} synthesized stellar parameters available in the literature for objects in the \kepler\ Input Catalog. For the M dwarfs, measurements primarily come from \citet{Muirhead2012} and \citet{Dressing2013}.

To update the planetary radii we use the planet-to-star radius ratio ($r/R_*$) and calculate $r$ using either our new stellar radius or that from \citet{Huber2014}.  To update planet equilibrium temperatures ($T_{eq}$), we assume that the planet radiates the same amount as heat as it receives, that heat is distributed isotropically, that the planet has an albedo of $a=0.3$, and that the star and planet radiate as blackbodies. This gives the familiar equation $T_{eq} = T_\sun \times (1-a)^{1/4}\sqrt{R_*/2d}$. The ratio between the planet-star distance and the stellar radius ($d/R_*$) is another directly measured transit parameter. We calculate $T_{eq}$ using either our new stellar effective temperature or that from \citet{Huber2014}.

\begin{figure}
\includegraphics[width=\linewidth]{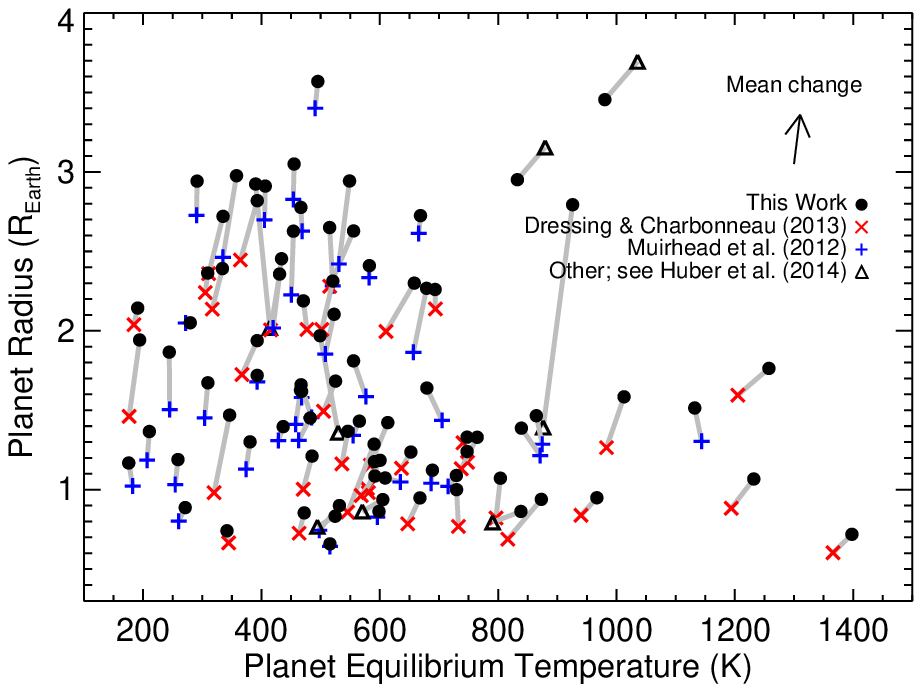}
\caption{Planet radius versus planet equilibrium temperature, when using stellar properties from \citet{Huber2014} or updated parameters from this work (filled circles). The \citet{Huber2014} catalog primarily draws M dwarf stellar parameters from \citet[][red crosses]{Dressing2013} and from \citet[][blue pluses]{Muirhead2012}, with small numbers of objects from other works (black open triangles). We use planet properties from the NExSci database (accessed 2014/06/10). Gray lines connect the previous and updated values for each planet.
\label{Fig:revised-params}}
\end{figure}

We present our updated stellar and planetary parameters in Table \ref{Tab:koi}. In Figure \ref{Fig:revised-params}, we show how the planet radii and equilibrium temperatures change when using our updated parameters. The difference in equilibrium temperature is largely negligible, but our new stellar radii have a significant effect on the radii of orbiting planets: the typical planet is $15\%$ larger with our stellar parameters than with those in \citet{Huber2014}.

\subsection{Comments on individual systems}

Two KOIs stand out because our new radii are smaller than those in \citet{Huber2014}. These are the candidate planets orbiting KOI 2715, for which the previous best stellar parameters come from the \kepler\ Input Catalog \citep{Brown2011}.

Three \kepler\ targets in our sample have previously received significant attention. KOI 961 (\kepler-42) hosts a suite of sub-Earth-sized planets and was analyzed by \citet{Muirhead2012a}, who inferred this star's properties by tying models to Barnard's star, which has a directly measured radius. While the temperature we estimate is nearly $200$K hotter than the temperature from their analysis, the two estimates are consistent ($3068\pm174$K versus $3254\pm110$K). The radii are also in very good agreement: $0.17\pm0.04\rsun$ in their analysis versus $0.19\pm0.04$ in this work. 

\citet{Johnson2012} presented KOI 254 (\kepler-45), an early M dwarf hosting a hot Jupiter, and found $R=0.55\pm0.11\rsun$ and $\teff=3820\pm90$, again consistent with our results of $R=0.58\pm0.03\rsun$ and $\teff=3870\pm80$.

KOI 571 (\kepler-186) was recently announced as hosting an Earth-sized planet in its habitable zone \citep[\kepler-186f,][]{Quintana2014}. The stellar parameters listed for this star in \citet{Huber2014} and in our Table \ref{Tab:koi} are from \citet{Muirhead2012}. \citet{Quintana2014} separately determined the radius for this star by finding the Dartmouth model best matching the mean stellar density, determined from transit photometry, and the metallicity and $\teff$ from \citet{Muirhead2012}. 
Their radius of $0.47\pm0.05\rsun$ is smaller than our estimate of $0.53\pm0.03\rsun$, but is consistent. We also revise the radius of \kepler-186f upward, from $1.02\rearth$ to $1.17\rearth$; a planet of this size is still likely to be rocky \citep[e.g.][]{Rogers2014}. The $\teff$ ($3624\pm80$) and luminosity ($0.048\pm0.008\lsun$) we infer for this object are also consistent with the properties from \citet{Quintana2014}, who estimated $\teff=3788\pm54$K and $L=0.041\pm0.009\lsun$. Therefore, our results support the conclusion that \kepler-186f is a rocky, habitable-zone planet.

\section{Summary}\label{Sec:summary}

We presented empirical calibrations for the effective temperatures, radii, and luminosities of cool dwarfs. We used 24 M dwarfs with interferometrically-measured parameters (25 for our radius calibration) to calibrate our relationships, which are based on EWs of \H-band spectral features. Our relationships are applicable to dwarfs with $3100<\teff \mathrm{(K)}<4800$, $0.2<R/\rsun<0.8$, and $-2.5<log L/\lsun < -0.5$. The standard deviations in the residuals of the best-fits are $73$K, $0.027\rsun$, and $0.049$ dex ($11\%$). From our bootstrap analysis, our luminosity calibration is the only one for which systematic error is important, but comparing temperature, radius, and luminosities indicates that  there may be additional sources of systematic uncertainty.  Our calibrations can be applied to stars without parallaxes and to non-flux calibrated spectra, and can be used very effectively for early M dwarfs with $3700<\teff \mathrm{(K)}<4000$, where the $\hind$ index used by \citet{Muirhead2012a} and \citet{Muirhead2014} to estimate temperature saturates. This is an important regime for understanding planets orbiting cool stars, because these are the late-type dwarfs with the greatest representation in \kepler. Figure \ref{Fig:teffrad} summarizes our results: we show the measured parameters for the interferometric stars on which the calibration is based, and the properties we infer for the MEarth and \kepler\ samples.

Our investigation of \H-band spectral features also revealed that the EWs we measure of features in our observed spectra that are not strongly dependent on metallicity -- in particular, \ion{Mg}{1} features -- show the best agreement with the EWs we measure from synthetic spectra. Conversely, the EWs we measure for features for which a metallicity dependence is apparent are the most discrepant.

We applied our calibrations to the MEarth sample of M dwarfs and validated the stellar parameters we infer by demonstrating that they display a clear relationship with $\mk$, which is an independent tracer of the stars' underlying physical properties. By comparing $\mk$ to inferred stellar luminosity, we identified $31$ candidate multiples. $18$ of the objects we identified in this manner are known binaries. We also used the luminosities we estimated to demonstrate that the \citet{Newton2014} metallicity calibration over-estimates the metallicities of late M dwarfs, and updated the metallicities of the sample using the calibrations from \citet{Mann2013} and \citet{Mann2014}.

Using spectra from \citet{Muirhead2014}, we applied our calibration to the cool stars from \kepler\ that host candidate planets. The temperatures that we find agree remarkably well with the temperatures reported in previous works, particularly given the different methods used, while our new stellar radii are larger. The largest discrepancy (median difference $0.09\rsun$) arises when we compare our radii to \citet{Dressing2013}, who fit photometry to Dartmouth models to estimate stellar parameters. The primary cause of the discrepancy is that at a given $\teff$, the interferometrically-measured radii are larger than those predicted by models by about $10\%$ \citep{Boyajian2012}, so the model-based radii from \citet{Dressing2013} are too small. The sub-solar metallicities they infer for the KOIs also contribute. Using our new stellar parameters, we updated the properties of the candidate planets, finding that the typical planet is larger than what one would calculate using the recent catalog from \citet{Huber2014} by $15\%$. The properties we infer for KOIs 961 (\kepler-42), KOI 254 (\kepler-45), and KOI 571 (\kepler-186) are consistent with the results from previously-published in-depth studies of those objects.

Our new calibrations have the benefit of being independent of stellar models. Because of the discrepancies between theoretical and observed stellar parameters, methods that rely on fixing stellar parameters to models will be subject to systematic errors. We note two important considerations that were discussed by \citet{Boyajian2012}. First, stellar radii for these low-mass stars generally are measured to be larger than predicted. Second, the effect of metallicity on radius does not appear to be as strong in the interferometric sample as is predicted by models, so metal-poor M dwarfs may be particularly misrepresented by such methods. We note, however, that we see some evidence of a metallicity dependence in the temperature-radius plane when considering the larger sample of stars to which we have applied our calibrations (see \S\ref{Sec:mann} and Figure \ref{Fig:teffrad}). Additional interferometric measurements -- particularly of mid-to-late M dwarfs and stars of extreme metallicity -- would improve both our empirical calibrations and our general understanding of the physical properties of low-mass stars.

\clearpage

\LongTables
\begin{deluxetable*}{r r r r l l r r r r r r}  
\tablecaption{\label{Tab:koi}Stellar parameters of cool Kepler Objects of Interest with high-fidelity spectra} \tabletypesize{\small} 
\setlength{\tabcolsep}{0.03in}
\tablecolumns{12}
\tablehead{
& 
& \multicolumn{3}{c}{\citet{Huber2014}}
& \multicolumn{3}{c}{This work} 
& \multicolumn{2}{c}{Literature\tablenotemark{c}}
& \multicolumn{2}{c}{This work} \\
\colhead{KOI Num.} 
&\colhead{\kepler\ ID}
& \colhead{$\teff$\tablenotemark{a}}
& \colhead{Radius\tablenotemark{a}}
& \colhead{Ref.\tablenotemark{b}}
& \colhead{$\teff$} 
& \colhead{Radius} 
& \colhead{Luminosity} 
& \colhead{$\mathrm{R}_\mathrm{p}$}
& \colhead{$\mathrm{T}_\mathrm{eq}$} 
& \colhead{$\mathrm{R}_\mathrm{p}$} 
& \colhead{$\mathrm{T}_\mathrm{eq}$}\\
& 
& \colhead{(K)} 
& \colhead{$R_\Sun$} 
&
 & \colhead{(K)} 
 & \colhead{$R_\Sun$} 
& \colhead{$\log{L/L_\Sun}$} 
& \colhead{$\rearth$}
& \colhead{K}  
& \colhead{$\rearth$} 
& \colhead{K} }
\startdata
 247.01&11852982&$3741$&$0.49$&SPE5&$3850\pm 75$&$0.556\pm0.027$&$-1.18\pm 0.06$&$ 1.85$&$ 507$&$ 2.10$&$ 522$\\
 250.01& 9757613&$3887$&$0.53$&SPE5&$3899\pm100$&$0.572\pm0.028$&$-1.10\pm 0.14$&$ 2.83$&$ 453$&$ 3.05$&$ 455$\\
 250.02& 9757613&$3887$&$0.53$&SPE5&$3899\pm100$&$0.572\pm0.028$&$-1.10\pm 0.14$&$ 2.70$&$ 405$&$ 2.91$&$ 406$\\
 250.03& 9757613&$3887$&$0.53$&SPE5&$3899\pm100$&$0.572\pm0.028$&$-1.10\pm 0.14$&$ 1.04$&$ 686$&$ 1.12$&$ 688$\\
 250.04& 9757613&$3887$&$0.53$&SPE5&$3899\pm100$&$0.572\pm0.028$&$-1.10\pm 0.14$&$ 2.73$&$ 290$&$ 2.94$&$ 291$\\
 251.01&10489206&$3810$&$0.52$&SPE5&$3827\pm 84$&$0.542\pm0.029$&$-1.20\pm 0.07$&$ 2.61$&$ 665$&$ 2.72$&$ 668$\\
 251.02&10489206&$3810$&$0.52$&SPE5&$3827\pm 84$&$0.542\pm0.029$&$-1.20\pm 0.07$&$ 0.83$&$ 595$&$ 0.86$&$ 598$\\
 254.01& 5794240&$3820$&$0.55$&SPE43&$3867\pm 82$&$0.578\pm0.028$&$-0.97\pm 0.10$&$10.23$&$ 691$&$10.71$&$ 700$\\
 255.01& 7021681&$3770$&$0.51$&SPE5&$4027\pm 81$&$0.616\pm0.027$&$-1.01\pm 0.12$&$ 2.46$&$ 334$&$ 2.98$&$ 357$\\
 255.02& 7021681&$3770$&$0.51$&SPE5&$4027\pm 81$&$0.616\pm0.027$&$-1.01\pm 0.12$&$ 0.74$&$ 497$&$ 0.90$&$ 531$\\
 314.01& 7603200&$3841$&$0.50$&SPE5&$3841\pm 73$&$0.512\pm0.027$&$-1.33\pm 0.05$&$ 1.58$&$ 467$&$ 1.62$&$ 467$\\
 314.02& 7603200&$3841$&$0.50$&SPE5&$3841\pm 73$&$0.512\pm0.027$&$-1.33\pm 0.05$&$ 1.68$&$ 392$&$ 1.72$&$ 392$\\
 314.03& 7603200&$3841$&$0.50$&SPE5&$3841\pm 73$&$0.512\pm0.027$&$-1.33\pm 0.05$&$ 0.64$&$ 515$&$ 0.66$&$ 515$\\
 463.01& 8845205&$3387$&$0.30$&SPE5&$3377\pm 79$&$0.372\pm0.028$&$-1.71\pm 0.06$&$ 1.50$&$ 244$&$ 1.87$&$ 244$\\
 478.01&10990886&$3744$&$0.50$&SPE5&$3727\pm 74$&$0.529\pm0.027$&$-1.27\pm 0.06$&$ 2.63$&$ 468$&$ 2.78$&$ 466$\\
 531.01&10395543&$4030$&$0.60$&SPE5&$4065\pm 76$&$0.630\pm0.028$&$-0.92\pm 0.07$&$ 3.40$&$ 490$&$ 3.57$&$ 495$\\
 571.01& 8120608&$3761$&$0.46$&SPE5&$3624\pm 79$&$0.525\pm0.029$&$-1.32\pm 0.07$&$ 1.44$&$ 705$&$ 1.64$&$ 679$\\
 571.02& 8120608&$3761$&$0.46$&SPE5&$3624\pm 79$&$0.525\pm0.029$&$-1.32\pm 0.07$&$ 1.59$&$ 576$&$ 1.81$&$ 555$\\
 571.03& 8120608&$3761$&$0.46$&SPE5&$3624\pm 79$&$0.525\pm0.029$&$-1.32\pm 0.07$&$ 1.21$&$ 870$&$ 1.39$&$ 839$\\
 571.04& 8120608&$3761$&$0.46$&SPE5&$3624\pm 79$&$0.525\pm0.029$&$-1.32\pm 0.07$&$ 1.45$&$ 484$&$ 1.66$&$ 466$\\
 571.05& 8120608&$3761$&$0.46$&SPE5&$3624\pm 79$&$0.525\pm0.029$&$-1.32\pm 0.07$&$ 1.02$&$ 182$&$ 1.17$&$ 175$\\
 596.01&10388286&$3678$&$0.47$&SPE5&$3635\pm 78$&$0.536\pm0.028$&$-1.35\pm 0.08$&$ 1.29$&$ 874$&$ 1.47$&$ 864$\\
 818.01& 4913852&$3721$&$0.52$&SPE5&$3723\pm 86$&$0.537\pm0.028$&$-1.36\pm 0.10$&$ 2.34$&$ 581$&$ 2.41$&$ 582$\\
 854.01& 6435936&$3593$&$0.47$&SPE5&$3694\pm 85$&$0.470\pm0.032$&$-1.12\pm 0.06$&$ 2.05$&$ 271$&$ 2.05$&$ 279$\\
 898.01& 7870390&$3893$&$0.52$&SPE5&$4025\pm 99$&$0.632\pm0.038$&$-0.71\pm 0.09$&$ 2.42$&$ 530$&$ 2.94$&$ 548$\\
 898.02& 7870390&$3893$&$0.52$&SPE5&$4025\pm 99$&$0.632\pm0.038$&$-0.71\pm 0.09$&$ 1.86$&$ 656$&$ 2.27$&$ 679$\\
 898.03& 7870390&$3893$&$0.52$&SPE5&$4025\pm 99$&$0.632\pm0.038$&$-0.71\pm 0.09$&$ 2.02$&$ 419$&$ 2.45$&$ 433$\\
 899.01& 7907423&$3568$&$0.42$&SPE5&$3636\pm 77$&$0.448\pm0.030$&$-1.28\pm 0.06$&$ 1.34$&$ 554$&$ 1.43$&$ 565$\\
 899.02& 7907423&$3568$&$0.42$&SPE5&$3636\pm 77$&$0.448\pm0.030$&$-1.28\pm 0.06$&$ 1.02$&$ 715$&$ 1.09$&$ 729$\\
 899.03& 7907423&$3568$&$0.42$&SPE5&$3636\pm 77$&$0.448\pm0.030$&$-1.28\pm 0.06$&$ 1.31$&$ 428$&$ 1.40$&$ 436$\\
 936.01& 9388479&$3581$&$0.44$&SPE5&$3544\pm 78$&$0.511\pm0.028$&$-1.40\pm 0.07$&$ 2.28$&$ 520$&$ 2.65$&$ 515$\\
 936.02& 9388479&$3581$&$0.44$&SPE5&$3544\pm 78$&$0.511\pm0.028$&$-1.40\pm 0.07$&$ 1.30$&$1143$&$ 1.51$&$1132$\\
 947.01& 9710326&$3750$&$0.46$&SPE5&$3780\pm 92$&$0.543\pm0.034$&$-1.24\pm 0.07$&$ 2.23$&$ 450$&$ 2.63$&$ 454$\\
 961.01& 8561063&$3068$&$0.17$&SPE41&$3254\pm106$&$0.185\pm0.043$&$-2.73\pm 0.07$&$ 0.86$&$ 570$&$ 0.94$&$ 605$\\
 961.02& 8561063&$3068$&$0.17$&SPE41&$3254\pm106$&$0.185\pm0.043$&$-2.73\pm 0.07$&$ 0.79$&$ 790$&$ 0.86$&$ 838$\\
 961.03& 8561063&$3068$&$0.17$&SPE41&$3254\pm106$&$0.185\pm0.043$&$-2.73\pm 0.07$&$ 0.77$&$ 494$&$ 0.83$&$ 524$\\
1085.01&10118816&$3939$&$0.52$&SPE5&$3777\pm 97$&$0.532\pm0.030$&$-1.07\pm 0.09$&$ 1.05$&$ 634$&$ 1.07$&$ 608$\\
1397.01& 9427402&$3957$&$0.54$&PHO2&$4104\pm107$&$0.624\pm0.032$&$-1.21\pm 0.16$&$ 2.01$&$ 501$&$ 2.31$&$ 520$\\
1408.01& 9150827&$4023$&$0.57$&SPE5&$4192\pm 84$&$0.631\pm0.029$&$-0.78\pm 0.07$&$ 1.31$&$ 462$&$ 1.45$&$ 482$\\
1408.02& 9150827&$4023$&$0.57$&SPE5&$4192\pm 84$&$0.631\pm0.029$&$-0.78\pm 0.07$&$ 0.80$&$ 260$&$ 0.89$&$ 271$\\
1422.01&11497958&$3517$&$0.37$&SPE5&$3580\pm 98$&$0.426\pm0.031$&$-1.50\pm 0.08$&$ 1.41$&$ 457$&$ 1.63$&$ 465$\\
1422.02&11497958&$3517$&$0.37$&SPE5&$3580\pm 98$&$0.426\pm0.031$&$-1.50\pm 0.08$&$ 1.45$&$ 303$&$ 1.67$&$ 309$\\
1422.03&11497958&$3517$&$0.37$&SPE5&$3580\pm 98$&$0.426\pm0.031$&$-1.50\pm 0.08$&$ 1.13$&$ 373$&$ 1.30$&$ 380$\\
1422.04&11497958&$3517$&$0.37$&SPE5&$3580\pm 98$&$0.426\pm0.031$&$-1.50\pm 0.08$&$ 1.19$&$ 206$&$ 1.37$&$ 210$\\
1422.05&11497958&$3517$&$0.37$&SPE5&$3580\pm 98$&$0.426\pm0.031$&$-1.50\pm 0.08$&$ 1.03$&$ 254$&$ 1.19$&$ 258$\\
1649.01&11337141&$3767$&$0.48$&PHO2&$3833\pm102$&$0.574\pm0.030$&$-1.29\pm 0.13$&$ 0.98$&$ 580$&$ 1.18$&$ 590$\\
1681.01& 5531953&$3608$&$0.40$&PHO2&$3722\pm132$&$0.483\pm0.033$&$-1.46\pm 0.09$&$ 1.00$&$ 470$&$ 1.21$&$ 485$\\
1681.02& 5531953&$3608$&$0.40$&PHO2&$3722\pm132$&$0.483\pm0.033$&$-1.46\pm 0.09$&$ 0.88$&$1193$&$ 1.07$&$1231$\\
1681.03& 5531953&$3608$&$0.40$&PHO2&$3722\pm132$&$0.483\pm0.033$&$-1.46\pm 0.09$&$ 0.78$&$ 647$&$ 0.95$&$ 667$\\
1702.01& 7304449&$3304$&$0.26$&PHO2&$3334\pm 99$&$0.339\pm0.031$&$-1.91\pm 0.08$&$ 0.82$&$ 796$&$ 1.07$&$ 803$\\
1843.01& 5080636&$3584$&$0.45$&PHO2&$3650\pm 92$&$0.529\pm0.031$&$-1.40\pm 0.09$&$ 1.16$&$ 536$&$ 1.37$&$ 546$\\
1843.02& 5080636&$3584$&$0.45$&PHO2&$3650\pm 92$&$0.529\pm0.031$&$-1.40\pm 0.09$&$ 0.73$&$ 463$&$ 0.85$&$ 472$\\
1867.01& 8167996&$3799$&$0.49$&PHO2&$3938\pm112$&$0.578\pm0.029$&$-1.20\pm 0.12$&$ 1.13$&$ 737$&$ 1.33$&$ 764$\\
1867.02& 8167996&$3799$&$0.49$&PHO2&$3938\pm112$&$0.578\pm0.029$&$-1.20\pm 0.12$&$ 2.01$&$ 415$&$ 2.36$&$ 430$\\
1867.03& 8167996&$3799$&$0.49$&PHO2&$3938\pm112$&$0.578\pm0.029$&$-1.20\pm 0.12$&$ 1.01$&$ 579$&$ 1.18$&$ 600$\\
1868.01& 6773862&$3950$&$0.56$&PHO2&$4163\pm133$&$0.627\pm0.035$&$-0.94\pm 0.12$&$ 2.14$&$ 316$&$ 2.39$&$ 334$\\
1902.01& 5809954&$3763$&$0.46$&PHO16&$3737\pm114$&$0.490\pm0.032$&$-1.36\pm 0.12$&$18.43$&$ 172$&$19.77$&$ 171$\\
1907.01& 7094486&$3901$&$0.54$&PHO2&$3851\pm109$&$0.591\pm0.033$&$-1.24\pm 0.17$&$ 2.01$&$ 476$&$ 2.19$&$ 470$\\
2006.01&10525027&$3809$&$0.46$&PHO2&$3792\pm 93$&$0.592\pm0.030$&$-1.24\pm 0.10$&$ 0.77$&$ 732$&$ 1.00$&$ 729$\\
2036.01& 6382217&$3903$&$0.52$&PHO2&$4060\pm112$&$0.589\pm0.031$&$-1.03\pm 0.15$&$ 1.49$&$ 504$&$ 1.68$&$ 525$\\
2036.02& 6382217&$3903$&$0.52$&PHO2&$4060\pm112$&$0.589\pm0.031$&$-1.03\pm 0.15$&$ 0.96$&$ 568$&$ 1.09$&$ 591$\\
2057.01& 9573685&$3900$&$0.54$&PHO2&$3997\pm118$&$0.585\pm0.029$&$-1.02\pm 0.11$&$ 1.14$&$ 636$&$ 1.24$&$ 652$\\
2130.01& 2161536&$3972$&$0.56$&PHO2&$4251\pm130$&$0.635\pm0.037$&$-0.75\pm 0.09$&$ 1.72$&$ 367$&$ 1.94$&$ 392$\\
2191.01& 5601258&$3724$&$0.46$&PHO2&$3910\pm107$&$0.567\pm0.030$&$-1.24\pm 0.11$&$ 1.15$&$ 584$&$ 1.42$&$ 613$\\
2306.01& 6666233&$3878$&$0.52$&PHO2&$4029\pm 89$&$0.616\pm0.029$&$-0.92\pm 0.10$&$ 0.94$&$1546$&$ 1.11$&$1607$\\
2329.01&11192235&$3815$&$0.50$&PHO2&$3929\pm135$&$0.624\pm0.036$&$-1.32\pm 0.13$&$ 1.26$&$ 983$&$ 1.58$&$1012$\\
2347.01& 8235924&$3972$&$0.56$&PHO2&$4084\pm108$&$0.609\pm0.029$&$-1.25\pm 0.14$&$ 1.07$&$1637$&$ 1.16$&$1684$\\
2542.01& 6183511&$3339$&$0.29$&PHO2&$3417\pm113$&$0.344\pm0.041$&$-1.65\pm 0.11$&$ 0.60$&$1365$&$ 0.72$&$1397$\\
2650.01& 8890150&$3735$&$0.40$&PHO2&$4040\pm102$&$0.599\pm0.028$&$-1.12\pm 0.10$&$ 0.98$&$ 319$&$ 1.47$&$ 346$\\
2650.02& 8890150&$3735$&$0.40$&PHO2&$4040\pm102$&$0.599\pm0.028$&$-1.12\pm 0.10$&$ 0.86$&$ 545$&$ 1.29$&$ 590$\\
2662.01& 3426367&$3410$&$0.34$&PHO2&$3646\pm128$&$0.471\pm0.031$&$-1.30\pm 0.11$&$ 0.69$&$ 816$&$ 0.94$&$ 872$\\
2704.01& 9730163&$3327$&$0.19$&PHO54&$3134\pm102$&$0.274\pm0.034$&$-2.17\pm 0.10$&$ 2.02$&$ 414$&$ 2.92$&$ 390$\\
2704.02& 9730163&$3327$&$0.19$&PHO54&$3134\pm102$&$0.274\pm0.034$&$-2.17\pm 0.10$&$ 1.36$&$ 529$&$ 1.97$&$ 499$\\
2705.01&11453592&$3400$&$0.27$&PHO54&$3592\pm134$&$0.534\pm0.044$&$-1.52\pm 0.18$&$ 1.39$&$ 876$&$ 2.79$&$ 925$\\
2715.01& 9837661&$4385$&$0.71$&KIC0&$4150\pm129$&$0.660\pm0.029$&$-0.90\pm 0.22$&$ 6.83$&$ 588$&$ 6.38$&$ 557$\\
2715.02& 9837661&$4385$&$0.71$&KIC0&$4150\pm129$&$0.660\pm0.029$&$-0.90\pm 0.22$&$ 3.69$&$1035$&$ 3.45$&$ 980$\\
2715.03& 9837661&$4385$&$0.71$&KIC0&$4150\pm129$&$0.660\pm0.029$&$-0.90\pm 0.22$&$ 3.16$&$ 879$&$ 2.95$&$ 832$\\
2764.01&10073672&$3952$&$0.55$&PHO2&$4124\pm117$&$0.608\pm0.029$&$-1.02\pm 0.15$&$ 1.59$&$1204$&$ 1.76$&$1257$\\
2839.01& 6186964&$3900$&$0.54$&PHO2&$3935\pm136$&$0.555\pm0.031$&$-1.13\pm 0.14$&$ 1.29$&$ 740$&$ 1.33$&$ 747$\\
2845.01&10591855&$3954$&$0.55$&PHO2&$4066\pm146$&$0.618\pm0.031$&$-0.78\pm 0.22$&$ 0.84$&$ 940$&$ 0.95$&$ 966$\\
2926.01&10122538&$3903$&$0.52$&PHO2&$4208\pm157$&$0.603\pm0.034$&$-0.95\pm 0.16$&$ 2.28$&$ 515$&$ 2.63$&$ 555$\\
2926.02&10122538&$3903$&$0.52$&PHO2&$4208\pm157$&$0.603\pm0.034$&$-0.95\pm 0.16$&$ 2.00$&$ 610$&$ 2.30$&$ 658$\\
2926.03&10122538&$3903$&$0.52$&PHO2&$4208\pm157$&$0.603\pm0.034$&$-0.95\pm 0.16$&$ 2.45$&$ 364$&$ 2.82$&$ 392$\\
2926.04&10122538&$3903$&$0.52$&PHO2&$4208\pm157$&$0.603\pm0.034$&$-0.95\pm 0.16$&$ 2.36$&$ 310$&$ 2.72$&$ 335$\\
2992.01& 8509442&$3952$&$0.55$&PHO2&$4088\pm141$&$0.578\pm0.034$&$-0.90\pm 0.19$&$ 2.04$&$ 184$&$ 2.14$&$ 190$\\
3090.01& 6609270&$3854$&$0.53$&PHO2&$3850\pm128$&$0.560\pm0.030$&$-1.29\pm 0.16$&$ 1.17$&$ 748$&$ 1.24$&$ 747$\\
3090.02& 6609270&$3854$&$0.53$&PHO2&$3850\pm128$&$0.560\pm0.030$&$-1.29\pm 0.16$&$ 2.14$&$ 694$&$ 2.26$&$ 693$\\
3282.01&12066569&$3894$&$0.54$&PHO2&$3944\pm127$&$0.571\pm0.031$&$-1.23\pm 0.13$&$ 2.24$&$ 305$&$ 2.36$&$ 309$\\
3414.01& 6023859&$3900$&$0.54$&PHO2&$3834\pm159$&$0.563\pm0.039$&$-1.32\pm 0.16$&$18.48$&$ 239$&$19.37$&$ 235$\\
3749.01&11547869&$3311$&$0.22$&PHO2&$3362\pm112$&$0.348\pm0.039$&$-1.81\pm 0.10$&$ 8.17$&$ 314$&$12.62$&$ 319$\\
4252.01&10525049&$3842$&$0.53$&PHO2&$3809\pm117$&$0.586\pm0.033$&$-1.36\pm 0.22$&$ 0.67$&$ 344$&$ 0.74$&$ 341$\\
4427.01& 4172805&$3668$&$0.43$&PHO2&$4037\pm155$&$0.573\pm0.050$&$-1.20\pm 0.14$&$ 1.46$&$ 176$&$ 1.94$&$ 194$\\
\enddata
\tablenotetext{a}{Stellar parameter in catalog from \citet{Huber2014}}
\tablenotetext{b}{Reference for \citet{Huber2014} data. Data primarily come from: SPE5 = \citep{Muirhead2014}; PHO2 = \citep{Dressing2013}; SPE41=\citet{Muirhead2012}}
\tablenotetext{c}{Planet properties one infers using the stellar parameters from \citet{Huber2014} and the planet parameters from the NASA Exoplanet Archive}
\end{deluxetable*}

\acknowledgements

ERN was supported throughout this work by a National Science Foundation Graduate Research Fellowship and AWM by the Harlan J. Smith Fellowship from the University of Texas at Austin. The MEarth Team gratefully acknowledges funding from the David and Lucille Packard Fellowship for Science and Engineering (awarded to D.C.). This material is based upon work supported by the National Science Foundation under grants AST-0807690, AST-1109468, and AST-1004488 (Alan T. Waterman Award). This publication was made possible through the support of a grant from the John Templeton Foundation. The opinions expressed in this publication are those of the authors and do not necessarily reflect the views of the John Templeton Foundation. We thank Z. Berta-Thomspon, J. Dittmann, and P. Muirhead for helpful conversations and the anonymous referee for comments and suggestions that improved this paper. This work used observations from the Infrared Telescope Facility, which is operated by the University of Hawaii under Cooperative Agreement no. NNX-08AE38A with the National Aeronautics and Space Administration, Science Mission Directorate, Planetary Astronomy Program. This research has made extensive use of data products from the Two Micron All Sky Survey, which is a joint project of the University of Massachusetts and the Infrared Processing and Analysis Center / California Institute of Technology, funded by NASA and the NSF, NASAÕs Astrophysics Data System (ADS), and the SIMBAD database, operated at CDS, Strasbourg, France.

\end{document}